\documentclass[a4paper,reqno]{amsart}
\usepackage[dvips]{graphicx}
\usepackage[obeyspaces]{url}
\usepackage[english, swedish]{babel}
\usepackage[T1]{fontenc}
\author{Frank G. Borg}
\thanks{Jyväskylä Universitet, Karleby Universitetscenter Chydenius, POB 567, 67101-Karleby, Finland. Email: \url{borgbros@netti.fi}}
\begin{document}
\title{Relativistisk rapsodi} 
\maketitle
\begin{abstract}
Föreliggande essä behandlar den Speciella Relativitetsteorin med historiska kommentarer, samt diverse försök att modifiera teorin. Trots mer än hundra år på nacken finns det inget som tyder på att Relativitetsteorin kommer att ersättas inom nära framtid med någon bättre lokal teori för rumtiden.
\end{abstract} 
 
\tableofcontents

\part{Den speciella relativitetsteorin}

\section{Inledning}

Den speciella (SR) och allmänna relativitetsteorin\footnote{I början använde Einstein enbart beteckningen 'relativitetsprincip'. Planck införde 1906 beteckningen 'Relativtheorie' medan A. H. Bucherer lanserade 'Relativitätstheorie' som började nyttjas av Einstein 1907. Felix Klein föreslog 1910 utan framgång namnet 'Invariantentheorie'. Efter 1915 skiljde Einstein mellan den 'speciella' och den 'allmänna relativitetsteorin'.  Se J. J. Stachel, \textit{Einstein's miraculous years: Five papers that changed the face of physics} (Princeton U. P., 1998) s. 102.} (ART) utgör två av den modärna fysikens grundpelare jämsides med kvantmekaniken. Relativitetsteorins ställning är - för att följa den inslagna metaforiska linjen - till den grad grundmurad att man kanske sällan ägnar den någon större uppmärksamhet. Förvisso finns det en och annan som t ex hoppas kunna påvisa att Lorentz-symmetrin inte gäller till hundra procent\footnote{D. Mattingly, ''Modern tests of Lorentz invariance'' \url{gr-qc/0502097}; M. Pospelov \& M. Romalis, ''Lorentz invariance on trial,'' Physics Today 40-46 (July 2004); C. M. Will, ''Special relativity: A centennary perspective,'' \url{gr-qc/0504085}.  Will anmärker att ''On the 100th anniversary of special relativity, we see that the theory has been so thoroughly integrated into the fabric of modern physics that its validity is rarely challenged, except by cranks and crackpots. It is ironic then, that during the past several years, a vigorous theoretical and experimental effort has been launched, on an international scale, to find violations of special relativity. The motivation for this effort is not a desire to repudiate Einstein, but to look for evidence of new physics 'beyond' Einstein, such as apparent violations of Lorentz invariance that might result from models of quantum gravity. So far, special relativity has passed all these new high-precision tests, but the possibility of detecting a signature of quantum gravity, stringiness, or extra dimensions will keep this effort alive for some time to come''. Vi återkommer till frågan om 'bortom' Einstein-fysiken i senare avsnitt. Här kan man inflika att inom modärn fysik är det ju en inbyggd reflex att associera symmetrier med brutna symmetrier varför inte heller Lorentz kan undgå misstankar om 'symmetribrott'.}. SR utgör en så integrerad del av de ''kovarianta'' och eleganta fältteoriformalismerna att den kanske själv nästan hamnat ur sikte\footnote{Detta är ett förhållande som B. Greene tar upp i förordet till en ny utgåva av A. Einstein, \textit{The meaning of relativity} (Princeton U. P., 2005, första utgåvan 1922 baserad på föreläsningar vid Princeton universitetet 1921, sista uppl. utkom postumt 1956). ''As a professional physicist, it is easy to become inured to relativity. Whereas the equations of relativity were once startling statements fashioned within the language of mathematics, physicists how now written relativity into the very mathematical grammar of fundamental physics. Within this framework, properly formulated mathematical equations automatically take full account of relativity, and so by mastering a few mathematical rules one becomes technically fluent in Einstein's discoveries. Nevertheless, even though relativity has been systematized mathematically, the vast majority of physicists would say that they still don't 'feel relativity in their bones' (...) but I can also attest to the undiminished feeling of awe I experience each time I pay sufficient attention to the details hidden within mathematics streamlined for relativistic economy, and come face to face with the true meaning of relativity. Space and time form the very arena of reality. The seismic shift in this arena caused by relativity is nothing short of an upheavel in our basic conception of reality.'' (s. vii - viii.) (B.G. har slarvat en aning med förordet och lyckas stava namnen Riemann, Friedman och Schwarzschild fel samt kallar den senare för en ryss.) Einsteins framställning är elegant och koncis och i litterär klass med Diracs berömda \textit{The principles of quantum mechanics} (Oxford U. P. 1930).}. Emellertid, den revolution som Einstein startade i fysiken är långt från över. Rummets och tidens problematik kommer att fascinera människan så länge hon är kapabel att tänka. I denna essä som kretsar kring SR kommer vi att använda metrikkonventionen $(+ - - -)$  för Minkowski-metriken $g_{\mu \nu}$.

\section{\label{SYM1}Symmetri I}

Relativitetsteorin\footnote{A. Einstein, ''Zur Elektrodynamik bewegter Körper,'' Ann. d. Phys. 17, 549-560 (1905).} formulerades (principiellt) som ett svar på diskrepansen som hade uppkommit mellan Newtons mekanik och Maxwells elektrodynamik. Klassisk mekanik omfattade 'Galileis relativitetsprincip' som för två inertiala referensystem i relativ translatorisk rörelse postulerar transformationen (1D-fallet)

\begin{eqnarray}
\label{eq:GALa}
x^\star &=& x - vt \\
\label{eq:GALb}
t^\star &=& t
\end{eqnarray}

\begin{figure}
	\centering
		\includegraphics[scale=0.7]{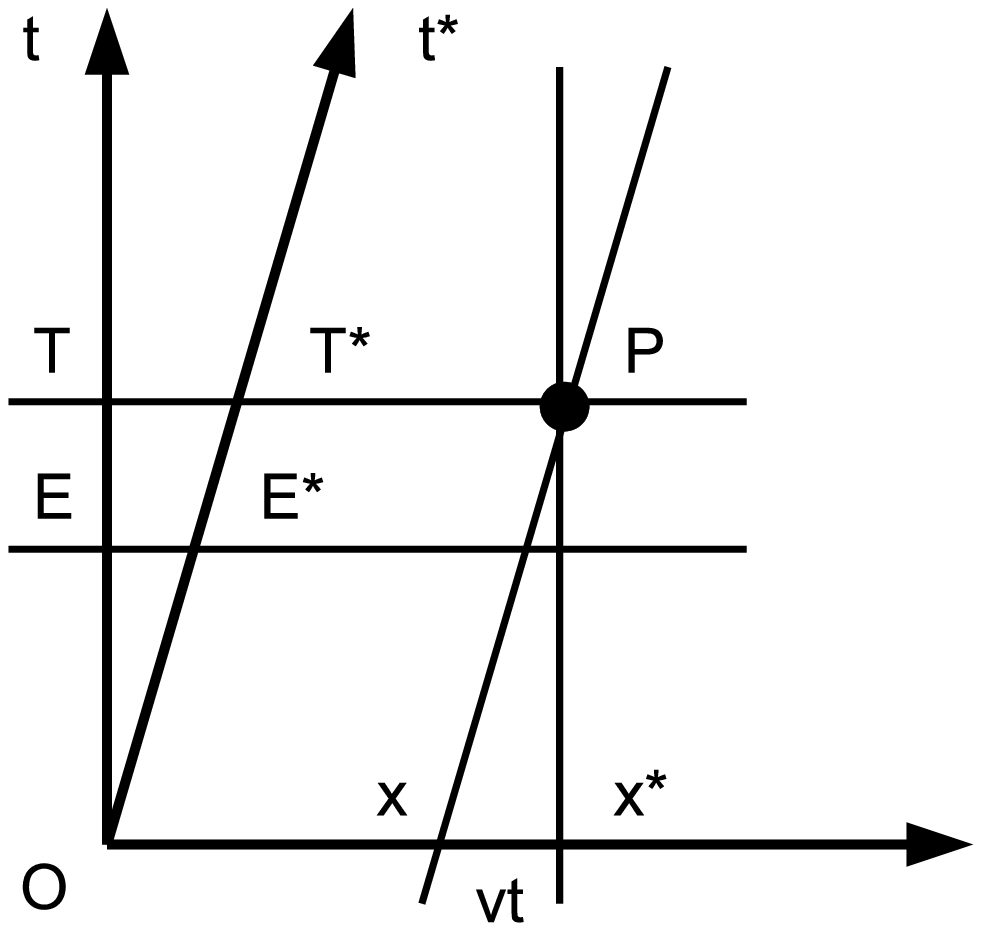}
	\caption{Galilei-transformationen}
	\label{fig:Fig1}
\end{figure}

Detta är vad det 'praktiska förnuftet' mer eller mindre föreskriver. Fig.\ref{fig:Fig1} ger en geometrisk beskrivning av transformationen (1). Måttlinjen (Eichkurve)  EE$^{\star}$  som träffar tidsaxlarna i punkterna E och E$^{\star}$ är införd för att vi skall få $t = t^{\star}$. Nämligen, för att erhålla de fysikaliska koordinatvärdena ($t$, $t^{\star}$) på axlarna $t$ och $t^{\star}$ skall vi dela med avståndena OE respektive OE$^{\star}$: $t$ = OT:OE; $t^{\star}$ = OT$^{\star}$:OE$^{\star}$. 

Einstein insåg att Galilei-transformationen (\ref{eq:GALa}, \ref{eq:GALb}) är inkonsistent med kravet på att ekvationen för ljusets utbredning (vi sätter härefter ljushastigheten i vakuum $c$ = 1)

\begin{equation}
\label{eq:LJUS}
t^2 - x^2 = 0
\end{equation}

\begin{figure}
	\centering
		\includegraphics[scale=0.7]{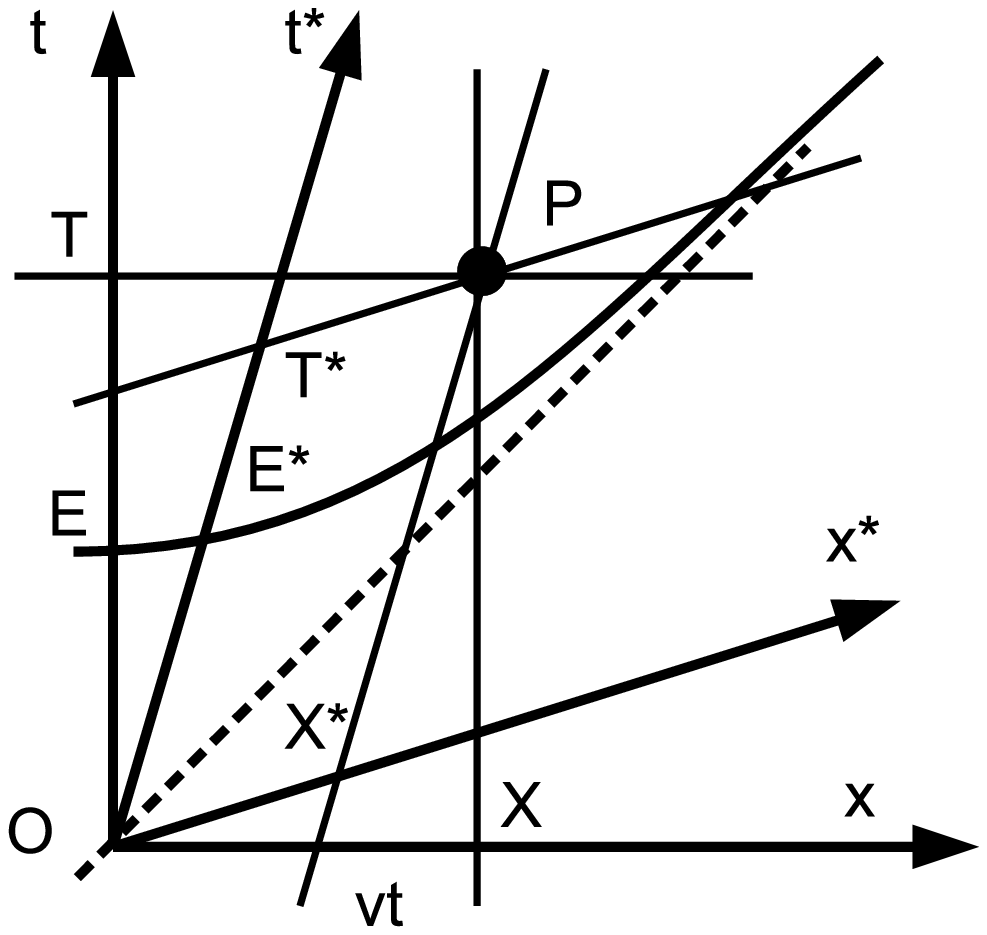}
	\caption{Lorentz-transformationen}
	\label{fig:Fig2}
\end{figure}

är invariant för inertialsystem och att den kräver en symmetrisk behandling av tiden och rummet i motsats till (\ref{eq:GALa}, \ref{eq:GALb}) och Fig.\ref{fig:Fig1}. I Fig.\ref{fig:Fig2} visas en symmetriserad transformation för $x$ och $t$.

Den streckade diagonalen är en del av ljuskonen (\ref{eq:LJUS}) och parabeln är en del av den nya måttlinjen

\begin{equation}
\label{eq:MATT}
t^2 - x^2 = 1.
\end{equation}

Den konventionella presentationen av den nya symmetriska transformationen brukar skrivas

\begin{eqnarray}
\label{eq:LORa}
x^{\star} &=& \gamma(v) (x - vt) \\
\label{eq:LORb}
t^{\star} &=& \gamma(v) (t - vx) \\
\label{eq:LORc} 
\gamma (v) &=& \frac{1}{\sqrt{1 - v^2}}.
\end{eqnarray}

Uttrycket för gamma-faktorn följer från invarianskravet och leder till Ekv.(\ref{eq:MATT}) för måttlinjen. Man kan också bestämma $\gamma$-faktorn genom att transformera tillbaka till ($x$, $t$)-systemet som använder samma ekvationer (\ref{eq:LORa})-(\ref{eq:LORb}) men med hastigheten $-v$. Transformationsformlerna följer också genom att observera att (\ref{eq:MATT}) kan parametriseras med hjälp av de hyperboliska funktionerna,

\begin{eqnarray}
\label{eq:HYPa}
t &=& \cosh(\phi_o)\\
\label{eq:HYPb}
x &=& \sinh(\phi_0)
\end{eqnarray}

En transformation (\ref{eq:LORa}, \ref{eq:LORb}) motsvaras då av en 'rotation'

\begin{eqnarray}
\nonumber
&&\phi_0 \rightarrow \phi_0 - \phi \quad \Rightarrow \\
\nonumber
&&x^{\star} = \sinh(\phi_0 - \phi) = \cosh(\phi) \left (\sinh(\phi_0) - \tanh(\phi) \cosh(\phi_0) \right) \\
\nonumber
&&= \gamma (x - v t).
\end{eqnarray}

Från uttrycket $v = \tanh(\phi)$ för hastigheten kan man enkelt härleda den relativistiska formen för additionsteoremet av hastigheter

\begin{equation}
\label{eq:ADD}
u \oplus v = \frac{u + v}{1 + uv}.
\end{equation}

Den geometriska tolkningen ovan går tillbaka på Hermann Minkwoski och har populariserats bl a av Max Born och Rolf Nevanlinna.\footnote{Max Born, \textit{Die Relativitätstheorie Einsteins} (3. uppl. Springer, 1922); Rolf Nevanlinna, \textit{Suhteellisuusteorian periaatteet} (WSOY, 1963). En annan elegant geometrisk härledning av transformationsformlerna baseras på den s.k. Bondi-faktorn, se H. Bondi, \textit{Relativity and common sense} (Dover 1980) - används t ex av M. Ludvigsen i \textit{General relativity. A geometric approach} (Cambridge U.P., 1999), kap. 3.} Minkowskis geometrisering (1908) var ett avgörande steg som sedan gjorde det möjligt för att Einstein att utveckla den allmänna relativitetsteorin.

\section{Egentid och gauge-teori}

I Minkowski-rum beräknas ett objekts 'egentid' $\tau$ genom ekvationen

\begin{equation}
\label{eq:TAU}
\tau = \int_{\Gamma} \sqrt{1 - v^2} \, dt,
\end{equation}

där $\Gamma$ betecknar dess bana i rumtiden. Givet en bana så är motsvarande förlopp av egentid en invariant storhet. Egentiden mellan två punkter A och B beror av vilken bana man följer vilket leder till den sk 'tvilling-paradoxen'. Beträffande 'tvilling-paradoxen' (eller 'klockparadoxen', 'Uhren-paradoxon') har jag faktiskt för många år sedan träffat på en matematiker som vid kaffebordet ihärdigt argumenterade för att dess 'lösning' kräver ART och ett beaktande av gravitationen.\footnote{R. Dugas har i \textit{History of mechanics} (Dover 1988) ett kort avsnitt som heter ''Pseudo-paradoxes in special relativity'' (495-7) där han beträffande tvilling-paradoxen anmärker att ''this paradox is so a pseudo-paradox: it is not in the field of validity of the Lorentz-Einstein theory'' utan ''outside the scope of special relativity''. Man kan undra vad som gett upphov till en tradition som ansett att accelererande rörelse ligger utanför SR? Att inertiala referenssystem har en särställning i SR betyder ju inte att SR inte kan hantera accelererande referens-system. Istället för en anomali kan 'tvilling-paradoxen' användas just för att framhäva en fundamental egenhet hos SR. Det kan anmärkas att A. Einstein också i ett skede 1918 sökte efter en sorts dynamisk förklaring till 'tvilling-paradoxen' baserad på ART. 'Tvilling-paradoxen' formulerad i termer av mänskliga resenärer går f.ö. tillbaka på en uppsats av Langevin 1911, se diskussionen i A. I. Miller, \textit{Albert Einstein's special theory of relativity. Emergence (1905) and early interpretations (1905-1911)} (Springer, 1998), s. 242-248.} Som synes är 'paradoxen' dock en direkt följd av vägberoendet av integralen (\ref{eq:TAU}). Man kan här uppfatta en analogi till mått-teorier (gauge theories). Faktiskt är det möjligt att använda just egentiden och dess vägberoende som ett sätt att belysa relativitetsteorins innebörd i analogi med dubbelspaltexperimentet i kvantmekaniken. Härifrån kan man då direkt gå vidare till den allmänna relativitetsteorin (ART) och förklara att den modifierar SR genom att mass-energi också påverkar egentiden. Själva begreppet egentid är tämligen enkelt att definiera populärt som den tid som en ideal medföljande klocka skulle ange.

Från (\ref{eq:TAU}) ser vi att maximal egen-tid mellan två punkter A och B i rumtid motsvarar en bana med långsammast möjliga förlopp ($v$ = 0). Om vi tänker oss att en sådan bana motsvarar en fri kropps rörelse så ligger det nära till hands att söka dylika rörelser som lösningar till en minimering av den negativa egentiden. Mer exakt så används inom SR formen

\begin{equation}
\label{eq:VERK}
S = - m \int \sqrt{1 - v^2}\, dt
\end{equation}

för verkan där $m$ betecknar objektets (vilo)massa. Ekv.(\ref{eq:VERK}) representerar den enklaste formen av en relativistisk invariant icke-trivial verkan. Genom verkan (\ref{eq:VERK}) kommer vi närmare en koppling mellan relativitetsteori och mått-teori eftersom verkan förekommer i kvantmekaniken genom fas-faktorn $\exp(iS/\hbar)$ (denna fas-faktor lyftes fram av Dirac\footnote{P. A. M. Dirac, ''The Lagrangian in quantum mechanics,'' Physikalische Zeitschrift der Sowjetunion, Band 3, Heft 1 (1933); se också ''On the analogy between classical and quantum mechanics,'' Rev. Mod. Phys. 17 (2 \& 3), 195-9 (1945). Feynmans doktorsavhandling godkändes 1942 och ges ut av L. M. Brown tillsammans med Diracs artikel i \textit{Feynman's thesis - a new approach to quantum mechanics} (World Scientific, 2005).} och utvecklades senare av R. P. Feynman till vägintegralformalismen).

\section{Einstein vs Poincaré}

Som bekant framlade Henri Poincaré Reltativitetsprincipen före Einstein och visade bl a att transformationen (\ref{eq:LORa})-(\ref{eq:LORb}), som han kallade för\footnote{Den tidigaste formen av (\ref{eq:LORa}) - (\ref{eq:LORb}) förmodas vara Woldemar Voigts framställning från 1887 som behandlade transformationerna som matematiskt knep vid lösningen av vågekvationerna.} 'Lorentz transformation', bildar en matematisk grupp. Ändå är det Einstein som av de flesta räknas som Relativtetsteorins upphovsman.\footnote{Enligt en modern myt, som nyligen t ex propagerats av en PSB-'dokumentär' och som visats bl a i Sverige (magasinet ''Vetenskapens värld'') och Finland, skulle Mileva Maric varit en dold medförfattare till Einsteins SR-artikel 1905. John Stachel ser sig nödsakad att i 2005 års upplaga av \textit{Einstein's miraculous years} (Princeton U. P., 2005) punktera denna historia som fötts genom en feltolkning av ett citat av Abraham Joffe. På ''Vetenskapens världs'' webbsida hittar man bl a påståendet att ''Vetenskapens värld lyfter också fram en rysk version av uppsatsen, förvarad i ett arkiv i Moskva, där man kan läsa dubbelnamnet Einstein-Maric.'' Stachels undersökningar visar att dokumentet i fråga är en kopia av en kopia av en passage ur en rysk populärvetenskaplig bok från 1962 som feltolkar ett citat av Joffe som ingenstans skrivit att någon artikel av Einstein skulle ha undertecknats med 'Einstein-Marty'. Det finns emellertid en passage där Joffe i en artikel 1955 berättar om de revolutionerande artiklarna från 1905 vars författare, ''an unknown person at that time, was a bureaucrat at the Patent Office in Bern, Einstein-Marity (Marity -  the maiden name of his first wife, which by Swiss custom is added to husbands family name.)'' Denna passage (där man utelämnar parentesen) utgör hela grundstenen för mytfabrikationen kring Mileva Maric som medförfattare till SR! Upphovskvinnan till myten är den serbiska författarinnan, Desanka Trubohovic-Gjuric, som skrivit en biografi över Mileva Maric översatt till tyska 1983.} En skillnad mellan Einstein och Poincaré har sagts vara att Poincaré var en 'fixare' som nöjde sig med att försöka 'lappa' ihop Lorentzs elektronteori. Enligt vetenskapshistorikern Peter Galison\footnote{P. L. Galison \& D. G. Burnett, 'Einstein, Poincaré \& modernity: a conversation'' (2003), \url{http://www.aip.org/history/einstein/essay-einsteins-time.htm}. R. Torretti å sin sida anser att Poincarés 'misslyckande' till stor del bestämdes av hans filosofiska syn: ''I am therefore inclined to attribute Poincaré's failure to another aspect of his philosophy, namely, his conventionalism''. (R. Torretti, \textit{Relativity and geometry} (Dover, 1996) s. 87.) Einstein utvecklade med tiden en tämligen sofistikerad syn på filosofi och epistemologi men med den brasklappen att samtidigt som forskaren behöver filosofisk-epistemologiska insikter har denne inte råd med att binda sig vid någon slutgiltig position enär i sista ändan är det verkligheten som bestämmer och den kan vi aldrig fullständigt känna. Vi kan inte i förväg veta vad vi har att vänta oss. Principer och 'ismer' skall inom forskningen vara vägledande men inte spika fast målet. Ph. Frank överraskades då Einstein 'övergav' den sorts positivism som synbart verkar ha varit utmärkande för skapandet av SR, medan M. Born t ex höll fast vid en deduktiv uppfattning av vetenskapen vilken Einstein inte heller omfattade. År 1919, då  Einstein i ett slag blev världsberömd för den stora allmänheten efter solförmörkelseexpeditionen vars mätningar kom att stöda hans teori, skrev han ett intressant inlägg för \textit{Berliner Tageblatt} (25.12.1919) vilket diskuteras och återges av A. Adam i ''Farewell to certitude: Einstein's novelty on induction and deduction, fallibism'' (J. for General Philosophy of Science 31, 19-37, 2000). I denna korta skrivelse framlägger Einstein fallibism-idén \textit{in nuce}, som senare skulle bli Karl Poppers varumärke (\textit{Logik der Forschung} (Springer, 1935)). Adam spekulerar i huruvida Popper (som då var 17 år och höll på att intressera sig för Einsteins teori) läst artikeln och inspirerats direkt av denna, en sak som dock kanske gör varken från eller till. Popper har ju ofta betonat att hans filosofi inspirerats av Einsteins relativitetsteori och att denna aktualiserat frågan om kunskapens gränser i och med att den 300 år gamla tanketraditionen (Newton) raserades; en tradition (som förstås inte sammanfaller med Newtons egna tankar) vilken betraktade mekaniken nästan som lika 'säker' som matematiken.  På ett ställe verkar A. Adam dock hugga i sten i sin uppsats. Nämligen, då Einstein i \textit{Tageblatt} skriver att en 'skarpsinnig forskare' ofta hamnar att samtidigt nyttja motstridiga teorier menar Adam att Einstein haft möjligen Maxwell eller Poincaré i tankarna, medan Einstein själv ''did not hold contradictory theories''. Snarare är det väl så att Einsteins \textit{annus mirabilis} var en strålande uppvisning i att just balansera med motsatser, hur han i en artikel lanserar hypotesen om ljuskvanta medan han i SR-artikeln återigen stöder sig på Maxwells fältteori. Den 'skarpsinniga forskaren' Einstein främst hade i tankarna får man förmoda var honom själv och med all rätt! Medan Einstein sannolikt antog att motsatserna skulle kunna förenas i en mer fullständig teori kan Niels Bohrs komplementraritetsfilosofi (\textit{Atomteori och naturbeskrivning} (Aldus, 1967)) ses som ett uttryck för antagandet att motsatserna karaktäriserar en intrinsisk dubbelhet hos naturen i relation till det epistemiska subjektet. Ett central medvetet 'motstridig' metod hos Bohr är t ex att analysera mätningar av kvantsystem i termer av 'klassiska' mätinstrument som ju 'egentligen' också borde beskrivas kvantmekaniskt.} representerade Poincaré en teknologisk syn på vetenskapen och en sorts ''reparative reason'' medan Einstein sökte efter en överenstämmelse mellan teori och verklighet. Thibault Damour\footnote{T. Damour, ''Poincaré, relativity, billiards and symmetry'' (\url{hep-th/0501168}).} radar upp en imponerande räcka upptäckter av Poincaré som kan räknas till Relativitetsteorin men ändå är hans slutsats att Poincaré inte kan sägas ha skapat Relativitetsteorin såsom en ny fundamental ram för fysiken. Poincaré t ex fortsatte efter 1905 att hålla fast vid Lorentzs distinktion mellan sann tid ('le temps vrai') och skenbar tid ('le temps apparent'). Enligt Poincarés synsätt skulle det således inte uppkomma någon verklig tidsskillnad i 'tvilling-paradoxen'. Det förefaller som om Poincaré tänkte sig att problemen i Lorentzs teori handlade om att fixa dynamiken och att han inte uppfattade saken som att Relativistetsteorin löste problemen genom att omgestalta kinematiken.\footnote{Man kan se ett snarlikt förhållningssätt i Hjalmar Tallquists 80-sidiga långa uppsats ''De fysikaliska förklaringarna af aberrationen'' (\textit{Festskrift Tillegnad Anders Donner} (Helsingfors 1915)). I uppsatsens avslutning tas faktiskt Relativistetsteorin upp men Tallquists omständiga genomgång av de 'fysikaliska' (dynamiska) förklaringarna till aberrationen visar att han vid denna tid inte helt övertygats om värdet av Einsteins kinematiska lösning. Liksom Poincaré och många andra hade Tallquist en sorts nostalgiskt förhållande till etern. I \textit{Naturvetenskapliga uppsatser} (Schildts, 1924) ägnar Tallquist ett kapitel åt ''Världseterns historia'' där han avslutningsvis utbrister, ''finnes det en världseter eller ej, så måste det uppriktiga svaret lyda, att det kan man ej veta''. Tallquists elev Gunnar Nordström blev tidigt medveten om Relatitivtetseorins revolutionära betydelse samt insåg genast att Minkowskis ''djärft matematiskt betraktelsesätt kastade ett fullkomligt nytt ljus'' över teorin, för att citera ur hans stilrena översikt ''Rum och tid enligt Einstein och Minkowski'' (Öfversigt af Finska Vetenskaps-Societetens Förhandlingar. LII. A, N.o 4, 1909-1910). Emellertid, den dynamiska tolkningstraditionen (Lorentz) lever vidare. Se t ex H. R. Brown \& O. Pooley, ''Minkowski space-time: a glorious non-entity,'' \url{physics/0403088}, som karaktäriserar den dynamiska/konstruktiva tolkningens kärnpunkt som så: ''What is definitive of this position is the idea that constructive explanation of 'kinematic' phenomena involves investigation of the details of the dynamics of the complex bodies that exemplify the kinematics'' (s. 11). Filosofen Harvey Brown har också på gång boken \textit{Dynamical relativity. Space time from a dynamical perspective} (Oxford U. P., 2005). A. Einstein betonade själv att SR var ofullständig såtillvida att entiteter som 'klockor' och 'mätstavar' inte visats motsvara lösningar till kovarianta ekvationer. I ett brev till A. Sommerfeld 14.1.1908 (citeras i Brown \& Pooley, s. 5) jämförde Einstein relationen mellan SR och en hypotetiskt mer fullständig teori som den mellan termodynamik och statistisk mekanik, en jämförelse som senare användes av J. Bell. Andra kända namn som efterlyst dynamisk förklaring till 'Lorentz-Fitzgerald-kontraktionen' är W. Pauli och A. S. Eddington. Antar man att SR bygger på en approximativ symmetri leds man också till frågan om hur den skall förklaras i termer av en 'djupare' struktur.} Därtill understryker Damour att Poincaré i sina pre-Einstein-1905 artiklar fortfarande laborerar med effekter av ordningen $O(v)$ och verkar vara tveksam att acceptera att Lorentz-transformationerna håller exakt.\footnote{S. Katzir, ''Poincaré's relativistic physics. Its origins and nature,'' Physics in Perspective 7, 268-292 (2005), går delvis emot Damours slutsatser. Enligt Kazir tog Poincaré Relativitetsprincipen på fullt allvar och bemödade sig om att visa att elektrodynamik och gravitationsteori kunde formuleras i samklang med denna. För Poincaré utgjorde således Relativitetsprincipen en sorts gränsvillkor, medan den för Einstein utgjorde en deduktiv utgångspunkt. Kazir betonar också att Poincaré uppfattade Lorentzs 'lokala tid' ('le temps apparent') som den tid som mäts av klockor, medan Lorentzs 'sanna tid' enbart är en konstruktion. Mitt intryck är dock att Damours argumentation väger tyngre beträffande inskränkningarna hos Poincarés Relativitetsprincip. Många fysiker och filosofer följde långt efter 1905 fortfarande de Lorentz-Poincaréska linjerna. Eino Kaila skrev ännu på 20-talet om 'etern' som han likställde med sorts elektromagnetisk bakgrundstrålning som erbjuder en global referensram, 'substraatti'. ''Niinpian kuin tämä myönnetään, ei Einsteinin radikaalista relativismia ajan ja avaruuden suhteen enää voidaan pitää pystyssä; ei voidaan silloin enää Einsteinin tavoin esim. väittää, että jokaisella taivaankappaleella on oma yksityinen aikansa, jonka ei tarvitse olla sama kuin muiden. Niin pitkälle kuin 'eetteri' ulottuu on myöskin aika yksi ja sama.'' (''Filosofisia huomautuksia relativiteettiteoriaan,'' Aika 14, 269-285 (1920). Förkortad version i \textit{Valitut teokset 1} (Otava, 1990).) Uppenbarligen verkar det som Kaila därmed skulle exempelvis ha avvisat 'tvillingparadoxen'. Senare i verket ''Einstein-Minkowskin invarianssiteoria'' (Ajatus 21, 5-121 (1958)) omfattar Kaila relativitetsteorin i hela sin fysikaliska vidd. En aning ironiskt är att Kaila noterar med tillfredsställelse att den äldre Einstein s.a.s. övergivit den Mach-inspirerade ''fenomenologiska fysiken''. Kailas egen filosofiska ståndpunkt verkade ju ursprungligen att ha lett honom vilse beträffande relativitetsteorins innebörd medan Einstein trots sina 'förvillelser' träffade rätt. Men liksom t ex Ph. Frank uppfattade Kaila kanske inte nyanserna i Einsteins tänkande och metoder utan försökte pressa in honom i trånga filosofiska fack. För en översikt av Einstein som filosof som betonar kontiuniteten i hans tänkande se D. A. Howard, ''Albert Einstein, philosophy of science,'' \textit{Standford encyclopedia of philosophy} (2004), \url{http://plato.stanford.edu/entries/einstein-philscience}. Howard anmärker att de som uppfattade 1905 års Einstein som Machisk fullblodspositivist glömmer att han med sina undersökningar kring Brownsk rörelse samma år sökte bevis på atomernas existens, i direkt motsats till Machs dåvarande uppfattning.} Poincaré dog 1912 utan att någonsin verkat ha omfattat Relativitetsteorins djupgående betydelse såsom vi förstår (?) den idag. Ändå har Damour fog för påståendet att det hade, ifall ett nobelpris ha utdelats för Relatitivtetseorin före 1912, varit rättvist att dela det mellan Einstein, Lorentz och Poincaré.\footnote{Två av Poincarés centrala artiklar om relativitetsteori, ''Sur la dynamique de l'électron,'' C. R. Acad. Sci., 140, 1504-1508 (1905); ''Sur la dynamique de l'électron,'' Rend. Circ. Math. Palermo 21, 29-175, (1906), kan hämtas från sajten ''Historic papers'', \url{
http://home.tiscali.nl/physis/HistoricPaper/Historic Papers.html}. På arkivet hittas också Einsteins centrala artiklar, samt artiklar av Minkowski, Laue,  m.fl. Två månader före sin död skrev Poincaré essän ''L'espace et le temps'' (ingår i \textit{Dernières pensées} (Flammarion, 1913)) som avslutas med de ofta citerade orden: ''Ajourd'hui certains physiciens veulent adopter une convention nouvelle. Ce n'est pas qu'ils soient contraints; ils jugent cette convention nouvelle plus commode, voilà tout; et ceux qui ne sont pas de cet avis pouvent légimitimement conserver l'ancienne pour ne pas troubler leurs vieilles habitudes. Je crois, entre nous, que c'est ce qu'ils feront encore longtemps.''}

\section{Konstant acceleration}

Ett viktigt specialfall inom den relativistiska kinematiken är konstant acceleration. Att ett objekt erfar konstant acceleration $a$ (i $x$-riktningen) i ett Minkowski-rum innebär att den under ett egentidsintervall $d\tau$ erfar en hastighetsökning $a \, d\tau$ jämfört med ett medföljande referenssytem. Använder man additionsteoremet (\ref{eq:ADD}) får vi för hastighetsökningen i 'laboratoriereferenssystemet'

\begin{equation}
\label{eq:ACC1}
dv = \frac{v + a \, d\tau}{1 + v a \, d\tau} - v \approx (1 - v^2) a \, d\tau = (1 - v^2)^{3/2} a \, dt.
\end{equation}

Detta kan skrivas som

\begin{equation}
\label{eq:ACC2}
\frac{d}{dt} \left ( \frac{dx}{d\tau}  \right) = a.
\end{equation}

Ekvationen (\ref{eq:ACC2}) kan tolkas som ekvationen för ett objekt som påverkas av en konstant kraft $F = ma$, varför den relativistiska generaliseringen av Newtons lag blir\footnote{M. Planck torde ha varit den förste som skrev ned denna form av relativistisk generalisering av Newtons ekvation i fallet med Lorentz-kraften (''Das Prinzip der Relativität und die Grundgleichungen der Mechanik,'' Verhandlungen der Deutschen Physikalischen Gesellschaft 4, 136-141, 1906). I artikeln 1905 hade Einstein ännu inte utvecklat en relativistisk mekanik. } (vilomassan betecknas med $m$)

\begin{equation}
\label{eq:DYN1}
m \frac{du}{dt} = \frac{d}{dt} \left( \frac{du}{d\tau} \right) = F
\end{equation}

med användning av den relativistiska definitionen av hastighet,  $u = dx/d\tau$. Kraften $F$ i (3) är den kraft som erfars av objektet visavis ett temporärt medföljande inertialt referenssytem. Inför vi den relativistiska impulsen $p = mu$ kan (\ref{eq:DYN1}) skrivas på den välbekanta formen $dp/dt = F$. Eftersom vi har identiskt

\begin{equation}
\label{eq:HYP2}
\left(\frac{dt}{d\tau}\right)^2 - \left(\frac{dx}{d\tau}\right)^2 = 1
\end{equation}

kan vi nyttja parametriseringen

\begin{eqnarray}
\label{eq:HYP3a}
\frac{dt}{d\tau} &=& \cosh(\phi(\tau)) \\
\label{eq:HYP3b}
\frac{dx}{d\tau} &=& \sinh(\phi(\tau)) 
\end{eqnarray}

som insatt i (\ref{eq:DYN1}) leder till $d\phi/d\tau = a$; dvs, den konstanta accelerationen har som lösning den hyperboliska banan

\begin{eqnarray}
\label{eq:HYP4a}
x(\tau) &=& \frac{1}{a} \left( \cosh(a \tau) - 1 \right) + x_0 \\
\label{eq:HYP4b}
t(\tau) &=& \frac{1}{a} \sinh(a \tau)
\end{eqnarray}

som ligger på en hyperbel $\left(x - x_0 + 1/a \right)^2 - t^2 = 1/a^2$.  Definierar vi energin $E$ sedvanligt som 'kraften gånger vägen' erhåller vi för den hyperboliska banan

\begin{equation}
\label{eq:ENE}
E(t) - E(0) = \int_0^t F \, dx = m \left( \cosh(a \tau) - 1 \right) = m \left (\frac{1}{\sqrt{1 - v^2}} - 1 \right) 
\end{equation}

vilket ger det relativistiska uttrycket för kinetisk energi som funktion av hastigheten $v$. Det relativistiska energibegreppet kan belysas genom att betrakta ett raketexempel som följer föregående härledning men i (\ref{eq:ACC1}) skall vi nyttja impulskonservering istället. Nämligen, antag att raketens massa vid tidpunken  $t$  är  $m$  och att dess propulsionssystem slungar ut en massa $\Delta m$ med hastigheten $w$ så att raketens massa (sett ur raketens referenssystem) ändras med  $dm$  under ett tidsintervall $dt$, då får vi för hastighetsändringen $dV$,

\begin{equation}
\label{eq:RAK1}
m\,dV + w \, dm = 0 \Rightarrow dV = - w \frac{dm}{m}, 
\end{equation}

som genom motsvarande resonemang som för (\ref{eq:ACC1}) (med  $a \, d\tau  = dV$) ger i 'laboratoriereferenssystemet'

\begin{equation}
\label{eq:RAK2}
dv = - \left(1 - v^2 \right) w \frac{dm}{m} \Rightarrow v = \frac{(m_0/m)^{2w} - 1}{(m_0/m)^{2w} + 1}  
\end{equation}

för hastigheten $v$, där $m_0$ betecknar raketens initialmassa då $v = 0$ (för $w \ll 1$ närmar sig den sista delen i (\ref{eq:RAK2}) K. Tsiolkovskys (1837-1935) klassiska raketekvation). En poäng här är att raketens mass-ändring $dm$ inte är lika med den massa $\Delta m$ som (accelereras i raketens motor och) slungas ut utan dessa är relaterade genom $dm = {\Delta m}/\sqrt{1 - v^2}$ vilket följer av energikonserveringen och (\ref{eq:ENE}). Einstein presenterade mass-energi formeln någon månad\footnote{A. Einstein, ''Ist die Trägheit eines Körpers von seinem Energieinhalt abhängig?,'' Ann. d. Phys. 18, 639-641 (1905). Einsteins härledning av (\ref{eq:ENE}) var begränsad till fallet med en elektrisk partikel i ett konstant elektriskt fält. Som sagt, 1905 hade han inte ännu tagit steget fullt ut till en relativistisk mekanik.} efter relativitetsartikeln fastän formeln (\ref{eq:ENE}) (för en elektron som accelererar i ett konstant elektriskt fält) redan ingår i relativitetsartikeln. Einstein var inte nöjd med 'beviset' för mass-energi formeln och sökte i flera år efter ett generellt bevis; hans sista publicerade bevisvariant är från 1946\footnote{M. Jammer ger en intressant översikt av olika ansatser för att grunda '$E = mc^2$' i \textit{Concepts of mass in contemporary physics \& philosophy} (Princeton U. P., 1999), kap. 3. Ovan underströks att $m$ betyder här 'vilomassan' av den anledningen att i litteraturen figurerar ett begrepp som 'relativistisk massa' som har vissa historiska anor men som är tämligen passé inom modärna framställningar av SR. Jammer diskuterar också debatten kring relativistisk massa. En färsk kritisk översikt om dess användning ges av G. Oas, \url{physics/0504110}. }. Numera vågar man kanske säga att gruppteorin har klarlagt massans roll i galileisk och relativistisk dynamik (se nedan).

\section{Geometri och fysik}

Fallet med konstant acceleration är också av intresse eftersom det leder till det icke-triviala problem att försöka konstruera ett global referens- och koordinatsystem för det accelererande objektet. Efter ekvivalensprincipens införande (Einstein 1907) så motsvarar detta uppgiften att konstruera ett globalt referenssystem i fallet med ett konstant homogent gravitationsfält. När man s.a.s. tittar i historiens bakspegel kan det förefalla som om utvecklingen av ART kunde ha tagit en helt annan fart ifall man först noggrannt studerat exemplet med konstant acceleration ur differentialgeometrisk och topologisk synvinkel. Detta är förstås ett anakronistiskt synsätt eftersom även dessa grenar av matematiken var delvis i sin linda. Även inom ART tog det t ex länge innan man förstod att skilja mellan äkta singulariteter och sådana som enbart berodde på 'olämpligt' val av koordinatsystem (såsom i fallet Schwarzschild-radien)\footnote{Naturen av Schwarzschild koordinat-singulariteten klarlades definitivt först av M. D. Kruskal i ''Maximal extension of Schwarzschild metric,'' Phys. Rev. 119, 1743-1745 (1960), även om t ex en lösning fanns redan implicit i ett arbete av A. S. Eddington från 1924. För en historisk notis om fallet se W. Rindler, \textit{Essential relativity} (2. ed. Springer, 1977), s. 150. Som ett tillägg kan man erinra sig att inom mått-teorier kan vi ha växelverkan fastän krökningen försvinner överallt utom i vissa singulära punkter. Aharonov-Bohm-effekten (Phys. Rev. 115, 485, 1959) är ett berömt exempelfall. Gravitationell effekt i rumtid med $R_{\mu \nu \kappa \lambda} = 0$ måste däremot rimligtvis vara utesluten.}. Einstein har själv påpekat att ett av de största hindren när han efter SR försökte utveckla ART var att han länge förbisåg skillnaden mellan koordinater som mer eller mindre godtyckliga matematiska konventioner och den intrinsiska geometrisk-fysikaliska innebörden.\footnote{J. Earman \& C. Glymour har följt upp Einsteins vingliga väg i ''Lost in the tensors: Einstein's struggles with the covariance principles 1912-1916,''  Stud. Hist. Phil. Sci. 9 (4), 251-278 (1978).} Einstein vägleddes i början av 'kovarians-principen' enligt vilken fysikens 'lagar' skall kunna formuleras på en koordinat-oberoende form. Emellertid, då Einstein i samarbete med Marcel Grossman undersökte möjligheten av en fältekvation av formen

\begin{equation}
\label{eq:GEN}
R_{\mu \nu} = \kappa T_{\mu \nu}
\end{equation}

som förknipper geometrin (Riemann-tensorn $R_{\mu \nu \kappa \lambda}$) med materien (energi-impuls-tensorn $T_{\mu \nu}$ ), så var de till en början fångna av ett gravt matematiskt missförstånd. Nämligen, de antog att $R_{\mu \nu} = 0$ implicerar att $R_{\mu \nu \kappa \lambda} = 0$ (ett förhållande som endast gäller i två dimensioner); dvs, att Riemann-tensorn försvinner identiskt. Härav skulle det följa att ingen växelverkan kan förekomma eftersom $T_{\mu \nu} = 0$  i rummet mellan materien (i vakuum) skulle implicera att därstädes är $R_{\mu \nu \kappa \lambda} = 0$. Einstein konstruerade följaktligen ett sinnrikt argument\footnote{Argumentet presenterades som en 'Bemerkung' till A. Einstein \& M. Grossmann, ''Entwurf einer verallgemeinerten Relativitätstheorie und einer Theorie der Gravitation,'' Zeitschrift für Mathematik und Physik 62, 225-261 (1913). Hålargumentet diskuteras av R. Torretti (1996, s. 162 - 168).} kallat för 'hålargumentet' för att bevisa att fysiken inte kan vara generellt kovariant! Även om Einstein korrigerade sitt misstag (själva argumentet är korrekt men frågan är vilka slutsatser man skall dra) råder det fortfarande delade meningar om vilken sorts fysikalisk innebörd man kan eventuellt tillmäta kravet på diffeomorfism-invarians.\footnote{Se t ex J. Barbour, ''Dynamics of pure shape, relativity and the problem of time'' \url{gr-qc/0309089} samt uppsatser i \textit{Physics meets philosophy at the Planck scale} (C. Callender \& N. Huggett, eds., Cambridge U. P. 2001). Att Einstein fick problem med differentialgeometrin är knappast att förvåna. Vid den aktuella tiden var teorin för mångfalder som sagt inte ännu kodifierad och höll på att formuleras av kanske främst H. Weyl, E. Cartan och O. Veblen. Weyls \textit{Die Idee der Riemannschen Fläche} (1913) är en av milstolparna i preciseringen av mångfaldsbegreppet. Med H. Whitneys ''Differentiable manifolds'' (Annals of Mathematics 37, 645-680, 1936) är den modärna syntesen ett faktum.} Problemet ansluter sig till frågan vad som menas med en punkt i rumtiden: givet två rumtids-mångfalder $M_1$ och $M_2$, kan vi identifiera punkter i $M_1$ med punkter i $M_2$? Einsteins lösning var ett tidigt exempel på insikten om att fysikaliskt meningsfulla storheter måste vara 'gauge-invarianta' (i detta fall för gruppen av diffeomorfismer).\footnote{C. Rovelli formulerar Einsteins slutsats som: ''Contrary to Newton and to Minkowski, there are no spacetime points where particles and fields live. There are no spacetime points at all. The Newtonian notions of space and time have disappeared''. (C. Rovelli: \textit{Quantum gravity} (Cambridge U. P., 2004) s. 74.)} Rumtidspunkterna har ingen objektiv realitet, endast 'koincidenser' mellan partiklar kan tillskrivas objektiv verklighet. Hänvisning till koincidenser leder emellertid småningom till nya svårigheter genom kvantmekaniken och kvantfältteorin; nämligen, begreppet lokalisering är problematisk.\footnote{H. Bacry, \textit{Localizability and space in quantum physics} (LNP 308, Springer 1988).}

\section{Symmetri II. No-interaction-theorem (NIT)}

Vad innebär det att en  teori är 'relativistiskt invariant'? Dagens fysiker har det i ryggmärgen att Lagrange-funktionerna skall vara relativistiskt invarianta. Kompletterar man Lorentz-transformationerna med translationer och rotationer erhåller vi Poincaré-gruppen $\mathcal P$. En märklig konsekvens som uppdagades under 60-talet är att invarians under Poincaré-gruppen för $N$-partikelsystem utesluter växelverkan; dvs, det enda fallet kompatibelt med 'manifest kovarians' är en samling fria partiklar. Detta resultat går under namnet \textit{No-Interaction-Theorem} (NIT) som kanske sällan tas upp i läroböcker.\footnote{Teoremet presenterades i originalversion av D. G. Currie, T. F. Jordan \& E. C. G. Sudarshan i ''Relativistic invariance and Hamiltonian theories of interacting particles,'' Rev. Mod. Phys. 35, 350-75 (1963), som ett par år senare utsträcktes till godtyckligt antal partiklar av H. Leytweyler (Nuovo Cimento 37, 556-67, 1965). En grundlig genomgång av den kanoniska versionen av teoremet och beviset ges i läroboken E. C. G. Sudarshan \& N. Mukunda, \textit{Classical dynamics - A modern perspective} (Wiley, 1974). Ett bevis som istället baseras på Lagrange-formulering presenterades av G. Marmo, N. Mukunda \& E. C. G. Sudarshan i ''Relativistic particle dynamics. Lagrangian proof of the no-interaction-theorem,'' Phys. Rev. D30, 2110-6 (1984). En elegant rendering av denna version återfinns i läroboken G. Esposito, G. Marmo \& E. C. G. Sudarshan, \textit{From classical to quantum mechanics} (Cambridge U. P., 2004), kap. 16.} För Galilei-gruppen däremot kan man lätt konstruera icke-trivial växelverkan på basen av translationsinvarianta och rotationssymmetriska interaktionspotentialer. Men detta lyckas inte i det relativistiska fallet ifall vi kräver att partiklarnas banor $(x(t), t)$ i Minkowski-rummet skall i likhet med koordinat-systemen även Lorentz-transformeras ('manifest koviarians', 'word-line condition'). Detta följer av att boost-generatorerna $K_i$ (som genererar Lorentz-transformationer) kommer att bero av dynamiken, de är inte rent kinematiska. Nämligen, eftersom Poincaré-gruppen föreskriver kommuteringsrelationerna för boost-generatorerna så sätter de därmed också villkor för dynamiken. Detta villkor visar sig vara så sträng att det endast kan uppfyllas för fria, icke-växelverkande partiklar. I Hamilton-dynamisk formulering\footnote{Elementa för Hamilton-mekanik, Poisson-klamrar mm återfinns i uppsatsen F. Borg, ''Symplektiska strukturer i fysiken'' (2000), \url{http://www.netti.fi/~borgbros/artiklar/sympl.pdf}.} utmynnar kravet på manifest kovarians i villkoret

\begin{equation}
\label{eq:MAN}
\{K^j,x^k\} = x^j \{H,x^k\}
\end{equation}

där $H$ betecknar Hamilton-funktionen. En nyckel-relation som tillsammans med (\ref{eq:MAN}) leder till NIT är $\{K^j,H\} = P^j$   där $P^j$ betecknar generatorerna för translationerna. För fria partiklar kan (\ref{eq:MAN}) enkelt uppfyllas med $K^j = x^j H$. En annan intressant omständighet är att Poisson-klammer mellan två boost-generatorer ger en rotation (rotationsgeneratorer betecknas med $J$),

\begin{equation}
\label{eq:ROT}
\{K^i,K^j\} = -\epsilon_{ijk} J^k 
\end{equation}

Detta leder till den s.k. Thomas-precessionen för t ex elektroner som 'kretsar' kring en atomkärna.\footnote{L. H. Thomas, ''The motion of the spinning electron,'' Nature 117 (2945), 514 (1926) (''Letters to the Editor''). H. Goldstein, \textit{Classical mechanics} (Addison-Wesley, 1989), 2. uppl., har kommenterat effekten: ''The spatial rotation resulting from the successive application of two parallel axes Lorentz transformations has been declared every as bit as paradoxical as the more frequently discussed apparent violations of common sense, such as the 'twin paradox''' (s. 287).} 
    	  
Hur är det då möjligt att vi trots NIT kan ha 'relativistisk' växelverkan? Den konventionella omvägen är att införa fält,\footnote{Såsom Peres noterar leder denna utväg i sin tur till nya problem .... ''the infinite number of dynamical field variables gives rise to new difficulties: divergent sums over states, far worse than those appearing when there is a finite number of continuous variables''. (A. Peres, \textit{Quantum theory - concepts and methods} (Kluwer, 1995) s. 256.)} av vilka det elektromagnetiska (EM) fältet $A_{\mu}$  (vektorpotentialen) är det primära exemplet (Faraday, Maxwell), som förmedlar växelverkan; d.v.s., partiklar växelverkar med fält istället för direkt med varandra genom potentialfunktioner. Emellertid finns det en intressant knorr när det gäller elektrodynamiken. Nämligen, det klassiska EM-fältet har inga egna frihetsgrader. Ekvationen (vi förutsätter bivillkoret $\partial_{\mu} A^{\mu} = 0$, s.k. 'Lorentz-gauge')

\begin{equation}
\label{eq:EM1}
\partial_{\mu} \partial^{\mu} A_{\nu} = 4 \pi J_{\nu}
\end{equation}

som förknippar fältet ($A$) med materien (strömmen $J$) kan omvändas enligt

\begin{equation}
\label{eq:EM2}
A_{\nu}(x) = \int G(x-y) J_{\nu}(y) \, dy
\end{equation}

genom en Greens funktion $G$ varigenom fältet $A$ kan elimineras ur teorin som formuleras direkt i term av partikeldata. Wheeler och Feynman påbörjade en dylik omformulering av elektrodynamiken ('absorber theory') under mitten av 1940-talet\footnote{J. A. Wheeler \& R. P. Feynman,  Rev. Mod. Phys. 17, 157 (1945); Rev. Mod.  Phys. 21, 425 (1949). Föregångare var H. Tetrode och A. D. Fokker. Det var försöket att kvantisera Wheeler-Feynman-teorin som i Feynmans dr:avhandling ledde fram till vägintegralmetoden, se L. M. Brown (2005).} vilken sedermera har utvecklats till 'action-at-a-distance electrodynamics' med avståndsväxelverkan. A. O. Barut har nyttjat den för en omformulering ('self-field' teorin) av kvant-elektrodynamiken liksom F. Hoyle och J. V. Narlikar vilka så sent som 1996 gav ut en lärobok i ämnet.\footnote{F. Hoyle \& J. V. Narlikar, \textit{Lectures on cosmology and action at a distance electrodynamics} (World Scientific, 1996).} Intresset för avståndsväxelverkan beror bl.a. på att den proberar fysikens teoretiska grunder samt att den ger nya synvinklar på renormaliseringsproblemet i kvantelektrodynamiken och självenergiproblemet i klassisk elektrodynamik.\footnote{M. Frisch diskuterar inkonsistensproblematiken i klassisk elektrodynamik i \textit{Inconsistency, asymmetry, and non-locality. A philosophical investigation of classical electrodynamics} (Oxford U. P., 2005). Maxwells ekvationer kombinerade med punkt-partiklar och Lorentz-kraften leder till problem, såsom divergerande själv-energi. Frischs tes är att det är omöjigt att kombinera Maxwell, Lorentz och punktpartiklar i en konsistent modell. Frischs poäng är emellertid att fysiker trots allt kommer väl till rätta med en inkonsistent teori genom att påföra extra villkor och kan därmed nyttja den med hög precision. För en matematisk studie av bl.a. Abraham- och Lorentz-modellen se H. Spohn, \textit{Dynamics of charged particles and their radiation field} (Cambridge U. P., 2004), \url{math-ph/9908024}.} (Själva formuleringen kan kanska verka 'ful' eftersom exempelvis gauge-teori-aspekten verkar att hamna i skymundan.) Problemkomplexet anknyter också till Bohms program för en ontologisk tolkning av kvantmekaniken.\footnote{D. Bohm \& B. J. Hiley, \textit{The undivided universe} (Routledge, 1993). Relativistisk invarians diskuteras i kap. 12.} Härvidlag tycks det råda en förvirring beträffande huruvida det finns en relativistisk Bohm-formulering eller ej. Om man som Bohm inför en partikel-ontologi för kvantmekaniken så är det klart att vi erhåller en icke-lokal teori (t ex EPR-situationen)\footnote{EPR efter Einstein, Podolsky \& Rosen vars beskrivning av 'entaglement-situationen' i kvantmekaniken är en klassiker (Phys. Rev. 47, 777-780, 1935). EPR spin-varianten lanserades av D. Bohm i \textit{Quantum theory} (Prentice-Hall, 1951).} vilken också var en av Bohms poänger. NIT-resultatet antyder också att det vore överraskande ifall formuleringen skulle leda till en manifest kovariant teori. Däremot är formuleringen relativistisk kovariant i den meningen att observablerna statistiskt sett uppfyller kovarianskravet. Således, relativistisk kovarians är därmed faktiskt mer relativ än absolut i Bohms formulering och teorin är öppen för situationer med ett 'överskridande' av relativitetsteorin och kvantmekaniken där s.a.s. icke-kovarians kan bli observerbar. I kvantfältteori möter man problemet med partiklar som 'skapas' och 'förintas' vilket verkar vara svårt att förena med en fundamental partikel-ontologi.\footnote{D. Dürr et al. (Phys. Rev. Lett. 93, 9, 090402, 2004) har dock byggt vidare på J. S. Bells arbeten (Phys. Rep. 137, 49, 1986) och formulerat kvantfältteori med Bohmska partikel-banor där partiklar skapas och förintas. H. Nikolic har i sin tur vidareutvecklat en kovariant kvantisering, utgående från s.k. De Donder-Weyl formalism, som han menar att naturligt bygger på deterministiska partikelbanor (\url{hep-th/0407228}).} I standardkvantmekaniken (J. v. Neumann)\footnote{J. v. Neumann, \textit{Mathematische Grundlagen der Quantenmechanik} (Springer, 1932).} förblir å andra sidan 'vågkollapsen' ett prekärt problem om man vill beskriva den med en kovariant dynamik.

\section{Symmetri III. Begreppet massa} 

Idag klassifierar vi partiklar i term av massa, spin och helicitet som bygger på representationer av Poincaré-gruppen. Den definitiva framställningen av detta klassifikationssystem är E. P. Wigners artikel från 1939\footnote{E. P. Wigner, ''On unitary representations of the inhomogenous Lorentz group,'' Annals of Mathematics, 40, 1, 149-204 (1939). Med 'inhomogenous Lorentz group' menas samma som Poincaré-gruppen.} om vilken Sternberg har anmärkt: ''It is difficult to overestimate the importance of this paper, which will certainly stand as one of the great intellectual achievements of our century. It has not only provided a framework for physical research for elementary particles, but has also had profound influence on the development of modern mathematics, in particular the theory of group representations.''\footnote{S. Sternberg, \textit{Group theory and physics} (Cambridge U. P., 1994), s. 149.} Wigners arbete var kulmineringen på en serie arbeten bl.a. av Majorana, Proca, Klein, Murray, Weyl, von Neumann och Dirac. Wigner ger speciellt erkännande åt Dirac: ''The subject of this paper was suggested to me as early as 1928 by P. A. M. Dirac who realized even at that date the connection of representations with quantum mechanical equations. I am greatly indebted to him also for many fruitful conservations about this subject, especially during the years 1934/35, the outgrowth of which the present papers is '' (Wigner 1939, s. 156.). Dirac\footnote{P. A. M. Dirac, ''Forms of relativistic dynamics,'' Rev. Mod. Phys. 21, 3, 392-399 (1949). Det kan f.ö. nämnas att det fanns en personlig länk mellan E. P. Wigner och Dirac genom att Dirac gifte sig med Wigners syster Margit 1937.} har också senare betonat att relativistisk dynamik ('klassisk' eller kvantum) handlar om representationer av Poincaré-gruppen; problemet gäller att finna nya dynamiska system vars observabler satisfierar Poincaré-algebrans kommuteringsrelationer. \textit{Classical dynamics} (1974)\footnote{Boken bygger f.ö. delvis på Mukundas föreläsningsanteckningar från V. Bargmanns kurs vid Princeton universitetet 1964-5.} av Sudarshan och Mukunda är ett prominent exempel på en tillämpning av detta program både för relativistisk och galileisk dynamik\footnote{I sammanhanget är det värt att nämna J. -M. Souriau som varit en pionjär i att förena symmetribegreppet med dynamik genom en generell symplektisk formulering av fysiken (\textit{Structure des systèmes dynamiques} (Dunod, 1970); eng. övers., Birkhäuser 1997). Souriau har också bidragit till teorin för geometrisk kvantisering (Souriau-Kostant teorin).} och följande utläggning är inspirerad av detta.

En kanonisk (symplektisk) realisering av Galilei-gruppen $\mathcal G$  (uppspänd av rotationer, rums- och tidstranslationer, samt Galilei-transformationer av typ Ekv. (\ref{eq:GALa})-(\ref{eq:GALb})) innebär att Lie-gruppen $\mathcal G$ är realiserad såsom kanoniska transformationer på ett fasrum $M$. Varje element $g$ i $\mathcal G$ motsvaras av en kanonisk transformation $\Phi_g: M \rightarrow M$ sådan att $\Phi_g \circ \Phi_h = \Phi_{gh}$; d.v.s. den definierar en representation av Galilei-gruppen $\mathcal G$  på gruppen av kanoniska transformationer, Kan$(M)$, som vi här kan definiera som gruppen av diffeomorfismer över $M$ vilka lämnar Poisson-klammern (PK) invariant. Denna representation inducerar vidare en representation av Lie-algebran $L(\mathcal G )$ på Lie-algebran av $\mbox{Kan}(M)$ vilken sammanfaller med mängden av Hamiltonska vektorfält på $M$; d.v.s., vektorfält $X_F$ som alstras av funktioner $F: M \rightarrow \mathbf{R}$ genom  $X_F f = \{f, F\}$ för en godtycklig reell funktion $f$  på $M$ ( $Xf$ betecknar derivatan av $f$  i riktningen  $X$: $Xf = df(X) = X^i \partial_if$). Givna två element $a$, $b$ i $L(\mathcal G)$ som motsvaras av generatorerna $F_a$ och $F_b$, då har vid det generella sambandet\footnote{Tilldelningen $a \rightarrow  F_a$  är vad man efter Souriau kallar för en \textit{momentum mapping} från $L(\mathcal G)$ till Kan$(M)$ där $\mathcal G$ betecknar systemets symmetrigrupp. Momentum-avbildningen generaliserar E. Noethers resultat om sambandet mellan symmetrier och invarianter och går tillbaka på S. Lie. (För en historisk notis se J. E. Marsden \& T. S. Ratiu, \textit{Introduction to mechanics and symmetry} (2. ed. Springer, 1999) s. 369-70.) }

\begin{equation}
\label{eq:COM}
\{F_a, F_b\} = F_{[a, b]} + C(a, b) 
\end{equation}

där $F_{[a,b]}$ betecknar en generator motsvarande Lie-produkten $[a, b]$ av $a$ och $b$ i $L(\mathcal G)$. $C(a, b)$ är en bilinjär funktion, en sorts 'integrationskonstant' och ett exempel på en \textit{kohomologi}. Genom omdefiniering av generatorer (genom att exempelvis addera konstanter) kan man försöka eliminera antalet 'integrationskonstanter'. Det intressanta är att för Galilei-gruppen $\mathcal G$  (i motsats till fallet för Poincaré-gruppen) återstår dock ett element som inte går att eliminera, nämligen $M$ i 

\begin{equation}
\label{eq:MAS}
\{P_i, G_j\} = - \delta_{ij} M.
\end{equation}

Här står $P_i$  för generatorn av translationer längs $x_i$-axeln, och $G_i$  för generatorn av Galilei-transformationen i dito riktning. Gruppen har alltså en icke-trivial kohomologi parametriserad av $M$ som kallas 'massa' (Souriau 1970).\footnote{V. Bargmann var antagligen den förste att studera (som ett exempel) representationer av Galilei-gruppen i ''On unitary ray representations of continuous groups,'' Annals of Mathematics 59, 1-46 (1954). Han visade bl.a. att Schrödinger-ekvationen motsvarar en Galilei-representation (för en enkel demonstration se Borg (2000)). En följd av att massan är ett neutralt element (= kommuterar med alla andra element) i Galilei-representationen är att massan  är 'superselekterad' (begrepp infört av Wick, Wightman och Wigner, Phys. Rev. 88, 101-105, 1952); d.v.s., ett tillstånd kan inte vara en superposition av tillstånd med olika massor. (I Wigner-Bargmann formuleringen manifesteras kohomologin i odeterminerade fasfaktorer för de unitära operatorer som representerar symmetrigruppen.) Enligt Primas betyder detta att ''the Galilei group gives the final explanation of the concept of the conservation of matter introduced into chemistry by Antoine Laurent de Lavoisier (1743-1794) and the law of definite proportions due to Joseph Louis Proust (1754-1826) and John Dalton (1766-1844).'' (H. Primas, \textit{Chemistry, quantum mechanics and reductionism}  (2. ed., Springer 1983) s. 73.) Primas uppger också att begreppet Galilei-transformation mellan referens-system infördes av Ph. Frank i ''Die Stellung des Relativitätsprinzips im System der Mechanik und Elektrodynamik,'' Sitzungsberichte d. math.-naturwiss. Kl. Akad. Wiss. Wien 118, 373-446 (1909). Begreppet 'referenssystem' i sin tur lanserades av Ludwig Lange 1885. L. Ricci har ett roligt inlägg i \textit{Nature} (vol. 434, s. 717,  7 april 2005) under rubriken ''Dante's insight into galilean invariance. The poet's vividly imagined flight unwittingly captures a physical law of motion''. Det handlar om Dante Alighieris \textit{Divina Commedia} (ca 1310) och raderna 115-117 där Dante beskriver en flygfärd på monstret Geryons rygg. Enligt Ricci visar beskrivningen att Dante intuitivt förstod att de fysikaliska betingelserna är likadana som i ett vilande referenssystem, förutom vindens påverkan. ... ''Dante intuitively grasped the concept of invariance but, unlike Galileo, he did not pursue this idea any further. Still, it seems he was well ahead of his time with regard to the views about laws of nature held in the Middle Ages''.  } Generatorn av tidstranslationer betecknas med $H$ (Hamilton-funktion) och dess enda icke-triviala PK är

\begin{equation}
\label{eq:MAS2}
\{G_i, H\} = P_i
\end{equation}

På basen av PK-relationerna kan man visa att\footnote{Som element av mängden av reella funktioner på fasrummet är det naturligt att konstruera dylika polynom. För en abstrakt Lie-algebra kan man också konstruera polynom genom att införa tensor-produkten på algebran (ett vektor-rum) och genom att identifiera $X \otimes Y - Y \otimes X$ med $[X, Y]$. Detta leder till s.k. 'enveloping algebras', se t ex A. O. Barut \& R. Raczka, \textit{Theory of group representations and applications} (World Scientific, 1986) s. 249-251.} 

\begin{equation}
\label{eq:CAS}
C = \mathbf{P}^2 - 2 M H
\end{equation}

är en Casimir-invariant (kommuterar med alla element). Ekv (\ref{eq:CAS}) kan skrivas på en mer bekant form

\begin{equation}
\label{eq:HAM}
H = \frac{\mathbf{P}^2}{2 M} + \mbox{konst.}
\end{equation}

som är den icke-relativistiska Hamilton-funktionen för en fri partikel med 'massan' $M$. Motsvarande undersökning kan göras för Poincaré-gruppen $\mathcal P$. Om man använder 4-vektorbeteckningen och sätter $P^0 = H$ (och använder metrikkonventionen $(+ - - - )$) så kan den första Casimir-invarianten skrivas

\begin{equation}
\label{eq:CAS1}
C_1 = P_{\mu} P^{\mu} = M^2
\end{equation}

som omvänt implicerar en Hamilton-funktion

\begin{equation}
\label{eq:HAM1}
H = \sqrt{\mathbf{P}^2 + M^2}.
\end{equation}

Jämför\footnote{P. Feyerabend har f.ö. gjort sig tolk för tesen att SR och klassisk mekanik är 'inkompatibla teorier' och att massbegreppet i den ena teorin inte kan jämföras med massbegreppet i den andra teorin (se Jammer 1999, s. 57). Emellertid, om man utgår från att båda teorierna måste hänföra sig till en och samma verklighet (de är s.a.s. representativa modeller) blir det rimligt att jämföra begreppen i dem. Matematiskt går det också att studera deformationer eller kontraktioner av Lie-algebran   $L(\mathcal G)$ på $L(\mathcal P)$ som visar hur vi en viss mening kan 'interpolera' mellan teorierna (V. Guillemin \& S. Sternberg, \textit{Symplectic techniques in physics} (Cambridge U. P., 1984) s. 114-116;  Barut \& Raczka (1986), s. 44-46). Inom vetenskapsfilosofin har man kanske ibland haft svårt att förstå fysikernas pragmatiska sätt att laborera med approximationer och övergångar från en teori till en annan beroende på 'validitetsområdet'. Men endel fysiker bidrar kanhända själva ibland till missförstånden genom att beskäftigt beskriva fysiken som en matematiskt exakt vetenskap eller som en 'teori om allting'.} vi den relativistiska versionen (\ref{eq:HAM1}) med (\ref{eq:HAM}) genom att utveckla den förra i serie av $\mathbf{P}^2$ då $\mathbf{P}^2 \ll M^2$   ser vi att för relativistisk symmetri är den additiva konstanten i (\ref{eq:HAM}) inte längre godtycklig utan ges av $M$ ('$M c^2$'). Således, relativistisk symmetri leder till att 'grundenerginivån' är fixerad till värdet $E_0 = Mc^2$. Märk även att medan massan är ett element i Galilei-algebran, är den enbart en konstant i det relativistiska fallet. 

Jammer (1999) återvänder upprepade gånger till 'mysteriet' varför ljusets hastighet $c$ figurerar i mass-energi formeln; bygger $E_0 = M c^2$  via någon sorts omväg på elektrodynamiken? Dock, SR handlar om en symmetrisering mellan rummet och tiden, varför det måste finnas en konstant '$c$' som förknippar $x$- och $t$-dimensionen och som gör att vi kan jämföra $\Delta x$ med $\Delta t$. Det är elektrodynamiken som ('anakronistiskt') ärver denna konstant genom att teorin formuleras såsom kompatibel med relativistisk kinematik. Galilei-symmetrin som har den enkla måttlinjen $t$ = konstant (se avsnitt \ref{SYM1}) undviker en sammanblandning av rum och tid, medan relativistisk symmetri har måttlinjen (\ref{eq:MATT}) vilken fodrar en konversionsfaktor mellan rum och tid. Mer korrekt vore kanske att säga att det på abstrakt nivå inte finns någon uppdelning i rum- och tidsdimension, det är vi som inför $c$ som gör denna uppdelning (i t ex meter vs sekund). Aristoteles som ryggade tillbaka inför att dividera kvantiteter med olika enheter (såsom längd/tid) kanske skulle ha känt sig mera hemmastadd i abstrakt relativistisk dynamik utan $c$.

\section{Tid och relativitet}

SR har självfallet djupt påverkat senare filosoferanden kring 'tidens väsen'. Den har inte löst några av tidens filosofiska problem, snarare förändrat perspektiven och gett upphov till nya frågor. Många har i SR funnit stöd för en statisk uppfattning ('tenseless time') kontra en dynamisk uppfattning ('tiden som flödar', 'tensed time') av tiden. Sett ur Minkowski-rum perspektiv är allting som hänt och händer redan fixerat i rumtiden. Einstein själv verkar ha varit benägen att tolka tiden statiskt, där upplevelsen av tiden som flödar är en sorts 'illusion'.\footnote{Ett av de klassiska Einstein-citaten är från ett brev till Michele Bessos familj efter denna nära väns bortgång 1955: ''Nun ist er mir auch mit dem Abschied von dieser sonderbaren Welt ein weinig vorausgegangen. Dies bedeutet nichts. Für uns glaubige Physiker hat die Scheidung zwischen Vergangenheit, Gegenwart und Zukunft nur die Bedeutung einer wenn auch hartnäckigen Illusion''.} Självfallet är inte SR nödvändig för att uppfatta verkligheten som fixerad i rumtiden, men efter SR och Minkowski går det inte längre att undvika att uppfatta verkligheten i term av rumtiden. Många filosofer och anhängare av statisk tid har inspirerats av SR, som t ex Michael Lockwood.\footnote{M. Lockwood, \textit{The labyrinth of time. Introducing the universe} (Oxford U. P., 2005). Lockwood gjorde sig bemärkt som vetenskapsfilosof genom \textit{Mind, brain and the quantum} (Oxford U. P., 1989).} Lockwood har tidigare tillsammans med David Deutsch utrett de logiska möjligheterna för tidsresor.\footnote{D. Deutsch, M. Lockwood, ''The quantum physics of time travel,'' Scientific American 270, 50-56 (1994).} Lockwood (2005) går igenom hela galleriet med tidens pil, entropi, statisk kontra dynamisk tid, kvantgravitation, tidresor, osv, (Wheeler-Feymann teorin avfärdas f.ö. som ''a dead duck'') men kanske de intressantaste anmärkningarna hittas mot slutet av boken. I fysiken (dess matematiska modeller) representeras 'tiden' ofta (åtminstone lokalt) som en parameter $t$, men det som intresserar oss här är hur den subjektiva tiden, upplevelsen av tiden som ett flöde, hänger ihop med den fysiska tiden $t$. Lockwood återvänder härvidlag till Everetts många-världars tolkning av kvantmekaniken.\footnote{H. Everett, '''Relative state' formulation of quantum mechanics,'' Rev. Mod. Phys. 29, 454-462 (1957); B. S. DeWitt, N. Graham, \textit{The many-worlds interpretation of quantum mechanics} (Princeton U. P., 1973).} Medvetandet förknippas med en speciell medvetande-bas $|\alpha_i \rangle$ i en Everett-dekomposition av universums vågfunktion $|\Phi \rangle$,  

\begin{equation}
\label{eq:EVE}
|\Phi \rangle = \sum |\alpha_i \rangle \otimes |\beta_i \rangle.
\end{equation}

Nu-upplevelsen (space-time-actuality) sägs uppstå då $|\alpha_i \rangle$ motsvarar en dekoherensbas, vilket hos Everett antagligen motsvarar de s.k. minnes-tillstånden (memory states), som har klassisk prägel. På neurofysiologisk nivå hänvisar Lockwood till teorierna om 'thalamocortical oscillations' och en form av 'cortical phase locking'.\footnote{Lockwood hänvisar bl.a. till R. Llinas, U. Ribary, ''Coherent 40-Hz oscillation characterizes dream state in humans,'' Proceedings of the National Academy of Sciences USA, 2078-2081 (1993).} Således, tidsmedvetandet kan uppdelas i 'specious presents' (subjektiva 'tidskvantum', en sorts minsta tidsupplevelse som uppfattas som en helhet, t ex observationen av ett stjärnfall på himlen) som betingas av att speciella neuroner i thalamus oskillerar i fas med varandra (en sorts koherent tillstånd). När dessa koherenta tillstånd bryter samman och ersätts av nya, stegar s.a.s. medvetandet fram i tiden (en tanke/förnimmelse ersätts av en annan tanke/förnimmelse). 

%\newpage

%\includegraphics[angle=90, scale=0.8]{Fig1.ps}

%\newpage

%\includegraphics[angle=90, scale=0.8]{Fig2.ps}

%\end{document}

\part{Kvantrum och SR-modifikationer}

\section{Dubbel speciell relativitet (DSR)}

På senare tid har det lanserats modifikationer av SR under den samlande beteckningen ''double special relativity'' (DSR)\footnote{G. Amelino-Camelia, ''Double special relativity,'' \url{gr-qc/0207049} v1; J. Magueijo \& L. Smolin, ''Lorentz invariance with an invariant energy scale,'' \url{hep-th/0112090}; ''Generalized Lorentz invariance with an invariant energy scale,'' \url{gr-qc/0207085}; ''Gravity's rainbow,'' \url{gr-qc/0305055}. Varför stanna vid 'dubbel-relativitet'? J. Kowalski-Glikman \& L. Smolin har infört en tredje invariant motsvarande den 'kosmologiska konstanten' $\Lambda$ i ''Triple special relativity'' (\url{hep-th/0406276}). DSR kan också läsas som 'deformed special relativity' som introducerades av F. Cardone \& R. Mignani 1998 och står för en teori där metriken $g_{\mu \nu}(E)$ beror på energin $E$ istället för att vara konstant som i SR (Cardone \& Mignani, \textit{Energy and Geometry: an introduction to deformed special relativity} (World Scientific, 2004); ''Energy-dependent metric for gravitation and the breakdown of local Lorentz invariance,'' Annales de la Fondation Louis de Broglie 27 no 3, 423-442, 2002). (Energi-beroende metrik diskuteras också av Magueijo \& Smolin, i ovan citerade \url{gr-qc/0305055}, men utan hänvisningar till Cardone \& Mignani.) Ytterligare en form av relativitetsteori lanserades av L. Nottali 1993 som han kallat 'scale-relativity' baserad på en sorts idé om fraktal rumtid men dess logik öppnade sig inte för undertecknad trots ett försök till läsning av hans detaljerade översikt ''Scale-relativity and quantization of the universe. I. Theoretical framework'' (Astronomy and Astrophysics, 327, 867-889, 1997). } där man uppfattar ljushastigheten $c$ som invariant men tillför också ytterligare en längd-invariant, alternativt mass- eller impuls-invariant. Motiveringen är att Planck-längden $L_P = \sqrt{G \hbar/c^3}$ ($\approx 10^{-35}$ m, eller  Planck-massan $M_P = \sqrt{\hbar c/G} \approx 10^{-8} \, \mbox{kg} \, \approx 10^{19}$ proton-massor), som karaktäriserar en hypotetisk kvantgravitationsregim, borde vara densamma för alla observatörer och inte påverkas av Lorentz-Fitzgerald kontraktionen: ''all observers agree that there is an invariant energy scale, which we take to be the Planck scale $E_P$'' (Magueijo \& Smolin, 2001). Denna motivering som sådan verkar ganska tunn; snarare handlar helt enkelt om antagandet att 'Planck-nivå' effekter fenomenologiskt förväntas yttra sig i en modifikation av t ex Lorentz-transformationerna. En ansats bygger på en ändring av den relativistiska 'dispersionsrelationen' (Magueijo \& Smolin 2002),

\begin{equation}
\label{eq:DSR1}
E^2 f_1(E, \lambda)^2 - \mathbf{p}^2 f_2(E, \lambda)^2 = m^2.
\end{equation}

$\lambda$ betecknar en karaktäristisk längdparameter som antas vara av storleksordningen $L_P$, $\lambda \approx L_P$. För $f_1 \neq f_2$ erhåller vi s.k. 'variable speed of light'-teorier (VSL) där 'hastigheten' $v = dE/dp$ för fotoner ($m$ = 0) kommer att bero av energin (eller frekvensen) vilket är av ett visst intresse inom astrofysiken; nämligen, en VSL-effekt skulle kunna skönjas som en frekvensberoende tidsfördröjning hos fotoner från gammakällor.\footnote{För en kartläggning av möjliga observationer av DSR-effekter se T. Jacobson et al., Phys. Rev. D67, 124011 (2003). J. Magueijo har väckt endel uppmärksamhet med sin bok \textit{Faster than light - the story of a scientific speculation} (Perseus Books, 2003) som innehåller en färgstark skildring av hans försök att lansera den variabla ljushastighetsteorin och hur han upplevt sig bli motarbetad. ''Though we get some glimpses here of theorists grappling with an elusive idea, too much of the story comes off as puerile'', anser G. Johnson i en recension i \textit{New York Times} (Feb. 9, 2003) med hänvisning till Magueijos angrepp på 'etablissemanget' . Magueijos bidrag ''A genuinly evolving universe'' i Wheeler-Festschriften J. D. Barrow et al., \textit{Science and the ultimate realtiy - Quantum theory, cosmology and complexity} (Cambridge U. P., 2004) s. 528-549, avslöjar en nästan manisk aversion mot tanken på universella oföränderliga lagar, härav tesen om 'muterande naturlagar' ... ''I am very fond of this view of the nature of physical law, primarily because I dislike the alternative current of thought: that there is 'a law', and that we shall know it; that we are close to the end of theoretical physics; that we may dream of a 'final theory'. Such mystical views are too 'lawyer-minded' for my taste'' (\textit{ibidem} s. 530). I bakgrunden verkar också vara en sorts klaustrofobisk känsla av att en värld med t ex konstant ljushastighet skulle i praktiken omöjliggöra interstellära rymdresor och möten med intressanta utomjordingar. Sant, för oss som växt upp med ET och rymdbarerna i \textit{Star Wars} och \textit{Liftarens guide till galaxen} så är det onekligen en trist tanke att vi för all framtid vore liksom instängda under en liten kupa i solsystemet. Magueijo spekulerar att man kunde resa med överljushastighet (lokalt är dock $v < c$) längs kosmiska strängar (\textit{ibidem} s. 546). Fysikern och författaren Alastair Reynolds har med sin \textit{Relevation Space}-rymdsaga (fyra böcker sedan 2000) dock visat att rymdepik är möjlig trots subluminal transport ifall man har möjlighet till lämplig kryoteknik, eller man kan hindra åldrandet.} I ett enkelt exempel på DSR modifieras Lorentz-transformationer $\Lambda$ verkande på momentum-rummet enligt

\begin{eqnarray}
\label{eq:DSR2a}
&& U^{-1}(p_0) \,  \Lambda \, U(p_0) \\
\label{eq:DSR2b}
&& U(p_0) = \exp(\lambda p_0 D) \\
\nonumber
&&\mbox{med dilatationsoperatorn}\\
\label{eq:DSR2c}
&& D = p_{\mu} \frac{\partial}{\partial p_{\mu}} 
\end{eqnarray}

Transformationen $U$ hänger ihop med funktionerna $f_1$ och $f_2$ i (\ref{eq:DSR1}) genom

\begin{equation}
\label{eq:DSR3}
U(p_0, \mathbf{p}) = \left(f_1(E, \lambda) p_0, f_2(E, \lambda) \mathbf{p} \right).
\end{equation}

Förutsatt att $|\lambda p_0| < 1$ (garanterar att $U(p_0)$ konvergerar) har vi

\begin{equation}
\label{eq:DSR4}
U(p_0) p_{\mu} = \frac{p_{\mu}}{1 - \lambda p_0}.
\end{equation}

Följaktligen får vi för en 'boost'-transformation längs $z$-axeln i $(p, E)$-rummet

\begin{eqnarray}
\label{eq:DSR4a}
\tilde{p}_0 &=& \frac{\gamma (p_0 - v p_z)}{1 - \lambda (1 - \gamma)p_0 - \lambda \gamma v p_z}  \\
\label{eq:DSR4b}
\tilde{p}_z &=& \frac{\gamma (p_z - v p_0)}{1 - \lambda (1 - \gamma)p_0 - \lambda \gamma v p_z}.
\end{eqnarray}

Magueijo och Smolin kallar (\ref{eq:DSR4a})-(\ref{eq:DSR4a}) för Fock-Lorentz transformationer eftersom de (i rumtids-version) redan lanserades av V. Fock\footnote{V. Fock, \textit{The theory of space-time and gravitation} (Pergamon Press, 1964).} i ett försök att modifiera Lorentz-transformationer för kosmologiska avstånd. DSR-teorier bygger på en modifiering av Lorentz-boost operatorerna som i föregående fall motsvarar (i energi-momentum rummet)

\begin{eqnarray}
\label{eq:DSR5a}
\tilde{K}^i &=& U^{-1}(p_0) K^i U(p_0) = K^i + \lambda p^i D \\
\label{eq:DSR5b}
K^i &=& p^i \frac{\partial}{\partial p_0} - p_0 \frac{\partial}{\partial p_i}. 
\end{eqnarray}

Enda kommutator som ändras i Poincaré-algebran är 

\begin{eqnarray}
\label{eq:DSR6a}
[\tilde{K}^i, P^j] &=& P^0 \delta_{ij} + \lambda P^i P^j\,, \\
\label{eq:DSR6b}
[\tilde{K}^i, P^0] &=& P^i + \lambda P^i P^0\,, 	
\end{eqnarray}

som reduceras till de traditionella formerna då  $\lambda = 0$. Ekv (\ref{eq:DSR5a})-(\ref{eq:DSR5b}) är ett enkelt exempel på en s.k. deformation av Poincaré-algebran som de senaste två decennierna studerats under namnet av $\kappa$-Poincaré algebran.\footnote{J. Lukierski, A. Nowicki \& H. Ruegg, ''New quantum Poincaré algebra and $\kappa$-deformed field theory,'' Phys. Lett. B293, 344 (1992). Polska forskare förefaller ha varit speciellt aktiva på detta område. För en översikt (Ph.D-avhandling) av ''$\kappa$-teorier'' se A. Agostini, ''Fields and symmetries in $\kappa$-Minkowski non-commutative space time,'' \url{hep-th/0312305}. En räcka av olika deformationer kan alstras via ekv (\ref{eq:DSR5a}) genom att modifiera $U$. Kraven på att Lie-algebra Jacobi-likheten måste satisfieras, samt att rotationsinvarians måste respekteras, begränsar väsentligen möjliga deformationer till $\kappa$-varianterna. Olika tillåtna deformationer kan visas motsvara olika val av koordinatsystem på ett \textit{de Sitter}-rum (J. Kowalski-Glikman \& S. Nowak, \url{hep-th/0204245}; J. Kowalski-Glikman, \url{hep-th/0207279}; F. Girelli \& E. R. Levine,  \url{gr-qc/0412079}). } För transformationerna (\ref{eq:DSR2a})-(\ref{eq:DSR2b}) erhåller vi följande invariant

\begin{equation}
\label{eq:DSR7}
m^2 = \frac{p^2}{\left(1 - \lambda p_0\right)^2} \quad (p^2 \equiv p_{\mu} p^{\mu} ),
\end{equation}

som definierar 'massan' $m$. Identifierar vi 'energin' enligt $E = p^0$ följer relationen

\begin{equation}
\label{eq:DSR8}
E = \frac{\gamma m}{1 + \lambda \gamma m}, 
\end{equation}

för kinetisk energi för en partikel genom att transformera m.h.a. (\ref{eq:DSR4a}) mellan partikelns referenssystem ($p_z = 0$) och 'laboratoriet'. Ekv (\ref{eq:DSR8}) innebär att energin begränsas genom $E < 1/\lambda \approx E_P$ (Planck-energin). En idé är att detta inför en energi cut-off som i princip kan eliminera divergenser i kvantfälteorier. En annan slutsats är att den modifierade Poincaré-symmetrin gäller endast för den 'sub-Planckiska' regimen. Detta kan verka motsägelsefullt eftersom makroskopiska kroppar antas bestå av partiklar med $E < E_P$ men ändå förmodas satisfiera de konventionella SR-transformationerna. Problemet med att massan begränsas av Planck-massan brukar hänvisas till som 'the soccer-ball problem' (Amelino-Camelia).  En tanke är att för sub-Planck partiklar måste vi också modifiera summeringsreglerna för impuls och energi för vilka Magueijo och Smolin (2002) föreslår en icke-linjär summering (betecknad $\oplus$)

\begin{equation}
\label{eq:DSR9}
\frac{p_i^{(tot)}}{1 - \frac{\lambda}{N} p_0^{(tot)}} = \sum_{k=1}^N \frac{p_i^{(k)}}{1 - \lambda p_0^{(k)}} \equiv p_i^{(1)} \oplus \dots \oplus p_i^{(N)}, 	
\end{equation}

där $p_i^{(tot)}$ betecknar energi-impulsen för en kropp sammansatt av $N$ partiklar. I transformations-formlerna (\ref{eq:DSR4a})-(\ref{eq:DSR4b}) för $p_i^{(tot)}$ ersätts $\lambda$ med  $\lambda/N$ vilka därför närmar sig Lorentz-transformationerna för stora $N$. Dessa egendomliga summerings- och transformationsregler som beror på hur man uppdelar en kropp i icke-sammansatta 'partiklar' antyder på principiella problem med denna typs DSR.\footnote{R. Alosio et al., ''A note on DSR-like approach to space-time,'' Phys. Lett. B610, 101-106 (2005). Här argumenteras  att ifall man försöker införa en invariant längd i rumtiden i likhet med fallet för momentum-rummet ovan hamnar vi att ''abandon the group structure of the translations and leave[s] a space-time structure where points with relative distances smaller or equal to the invariant scale cannot be unambigiously defined''. F. Girelli \& E. R. Levine \url{gr-qc/0412079} har gett en allmännare teoretisk bakgrund för Magueijo-Smolin momentum-additionen (\ref{eq:DSR9}). Enligt denna är (\ref{eq:DSR9}) att jämföra med relativistisk addition av hastigheter (\ref{eq:ADD}) som har att göra med en komposition av två relativa referenssystem.} H. Bacry\footnote{H. Bacry, ''The foundations of the Poincaré group and the validity of general relativity,'' Reports on Mathematical Physics 53 (3), 443-473 (2003). Kan nämnas att Bacry (enligt egen utsago) tydligen ofta diskuterat fysik med A. Connes som är känd som en pionjär inom icke-kommutativ geometri.} har utgått från $\kappa$-Poincaré-algebra med relationen

\begin{equation}
\label{eq:DSR10}
P_0 = 2 \kappa \sinh \left( \frac{\tilde{P}_0}{2 \kappa} \right).
\end{equation}

När parametern $\kappa \rightarrow \infty$ återfår vi $P_0 = \tilde{P}_0$. $P_0$ uppfattas som en klassisk additiv energi-parameter medan $\tilde{P}_0$ står för den fysikaliska energin. Således, en galaxs massa $M$, som består av $N$ stjärnor var och en med massan $m$, bestäms enligt Bacry av relationen

\begin{equation}
\label{eq:DSR10b}
\sinh \left( \frac{M}{2 \kappa} \right) = N \sinh \left( \frac{m}{2 \kappa}, \right).
\end{equation}

varav följer att $M < N m$; d.v.s., 'summan' är mindre än delarna. Genom att anta att $\kappa \approx 10^{11}$ solmassor (motsvarar en typisk galax) anser Bacry att man kan förklara och avfärda ''the dark matter affair'' även om tankegången inte verkar helt solklar.\footnote{Begreppet 'mörk materia' härrör sig från observationen att ifall man räknar ihop all materia som kan observeras i galaxer m.h.a. av optiska teleskop och radioteleskop så tyder stjärnornas snabba rörelser i galaxer på att det måste finnas ett betydande överskott av materia som inte 'syns' för att kunna förklara rörelserna (genom den mörka materiens gravitationella påverkan). Yttre delar av galaxer roterar snabbare än vad den skenbara massan enligt Keplers lagar skulle föreskriva. Hypotesen om mörk materie går tillbaka på F. Zwicky och dennes observationer under 1930-talet av galaxerna i Comaklustern. Einstein diskuterar metoden för att bestämma andelen mörk materia i \textit{The meaning of relativity}, appendix 1. Enligt uppskattningar består 90-95\% av universum av mörk materia och energi ('mörk energi' förmodas förklara universums accelererande expansion). VIRGOHI21 som upptäcktes i februari 2005 med radioteleskop (tack vare 'väte-signalen' på våglängden 21 cm) tros vara en galax helt bestående av mörk materia. Vätgasens dynamik implicerar närvaron av den 'osynliga' galaxen. Existensen av ett sådan mörk galax innebär rimligtvis en dödsstöt för de flesta 'bortförklarings-teorierna'. (För en populär översikt av 'mörk materia' se \url{http://en.wikipedia.org/wiki/Dark_matter}.) ''[T]he issue of dark matter arguable constitutes the most astrophysically interesting aspect of modern cosmology'' (J. A. Peacock, \textit{Cosmological physics} (Cambridge U. P., 1999) s. 368), men huruvida denna fråga har något att direkt göra med relativtetsteorins fundament, eller kan 'bortförklaras' med revisioner av denna, är ovisst. I sig är existensen av mörk materia inget att förvåna sig över ... ''It is not surprising that most of the matter in the universe should be dark: there is no reason why everything should shine'' ... (M. Rees, \textit{Perspectives in astrophysical cosmology} (Cambridge U. P., 1995) s. 33). Mörk materia som bara växelverkar via gravitationen (s.k. WIMP) vore självfallet aningen kufisk. Ett annat aktuellt fenomen som också satt fart på många relativisters fantasi är den s.k. Pioneer-anomalin (se t ex  J. G. Harnett \& F. J. Oliveira, ''Anomalous acceleration of the Pioneer spacecraft matches the acceleration predicted by Carmelian cosmology,'' \url{gr-qc/0504107}. Analys av bandata från rymdsonderna Pioneer-10 och 11 antyder på att de utsätts för en oförklarlig extra radiell acceleration mot solen av storleksordningen $(8.74 \pm 1.33) \times 10^{-10}$ m/s$^2$ (Anderson et al, Phys. Rev. Lett. 81, 2858-2861, 1998). (Igen hittas en tämligen bra populär översikt på Wikipedia, \url{http://en.wikipedia.org/wiki/Pioneer_anomaly}.) Man förmodar att 'anomalin' blir kännbar på ca 20 AU:s avstånd från solen  (AU = 'astronomical unit' = medelavståndet mellan solen och jorden). En sådan effekt skulle fodra en förklaring på super-Planckisk nivå istället för på sub-Planckisk nivå. Ett märkligt samband som lett till en del spekulerande (exv Hartnett \& Oliveira) är att den anomala accelerationen är av samma storleksordning som $H c \approx 6.9 \times 10^{-10}$ m/s$^2$ där $H$ betecknar Hubble-konstanten ($\approx 2.3 \times 10^{-18}$ s$^{-1}$).} I Bacrys version blir deformationerna nu istället endast märkbara i kosmologiska sammanhang för galaktiska system medan i Magueijo-Smolin-teorin effekterna gällde den sub-Planckska regimen. Det kan verka aning långsökt att modifiera SR för kosmologiska skalor eftersom gravitationen dominerar på denna nivå och vi borde övergå till ART. 

Ett centralt problem med deformerade Poincaré-algebran är att de formulerats för momentum-rummet. Det finns ingen automatisk koppling till Minkowski rumtiden. Detta hänger ihop med att det saknas enhetliga principer för fysikaliska tolkningar av de algebraiska schematan. Således, när vi tidigare hänvisade till 'hastigheten' $v = dE/dp$ i samband med (\ref{eq:DSR1}) så hade vi ingen grund för att identifiera detta uttryck med hastighet i den 'vanliga' meningen $v = dx/dt$.  Den kanske mest systematiska ansatsen på att förbinda momentumrummet och rumtiden baseras på teorin för kvantgrupper (icke-kommutativa Hopf-algebran) som upprättar en sorts dualitet mellan $\kappa$-Poincaré-algebra och $\kappa$-Minkowski rumtid.\footnote{För en definition av dessa termer se t ex Agostinis artikel \url{hep-th/0312305}. 90-talets kanske mesta kvantgruppsguru har varit Shahn Majid som gett en pedagogisk översikt och introduktion i \textit{Foundations of quantum group theory} (Cambridge U. P., 1995, pb ed. 2000). Hans millenium-artikel ''Quantum groups and non-commutative geometry,'' J. of Mathematical Physics 41 (6), 3892-3942 (2000), uttrycker stora förhoppningar på att kvantgrupper och icke-kommutativ geometri är den rätta vägen till fysikens innersta hemligheter. ''In fact, there are fundamental reasons why one needs noncommutative geometry for any theory that pretends to be a fundamental one. Since gravity and quantum theory both work extremely well in their separate domains, this comment refers mainly to a theory that might hope to unify the two. As a matter of fact I believe that, through noncommutative geometry, this 'Holy Grail' of theoretical physics may now be in sight.'' Liksom Penroses twistorer (se nedan) så inställer sig frågan huruvida den eleganta matematiken enbart häller samma gammla vin i nya läglar eller ifall den faktiskt kan inspirera till helt ny fysik.} En kanske mer 'fysikalisk' koppling har gjorts av Magueijo, Smolin, m. fl. Nämligen, utgångspunkten är att teorin skall återge planvågslösningen $\exp(i p_{\mu} x^{\mu}/\hbar)$ för fria partiklar och behålla uttrycket för den invarianta fasen $p_{\mu} x^{\mu}/\hbar$. Detta implicerar via (\ref{eq:DSR1}) en sorts dual energiberoende rumtidsmetrik av formen\footnote{D. Kimberly et al., \url{gr-qc/0303067}; J. Magueijo \& L. Smolin, \url{gr-qc/0305055}.}

\begin{equation}
\label{eq:DSR11}
ds^2 = \frac{dt^2}{f_1(E, \lambda)^2}-\frac{d\mathbf{x}^2}{f_2(E, \lambda)^2},
\end{equation} 

som motsvarar en dual avbildning till (\ref{eq:DSR3}) given genom

\begin{equation}
\label{eq:DSR12}
U(x) = \left( \frac{t}{f_1(E, \lambda)}, \frac{\mathbf{x}}{f_2(E, \lambda)} \right).
\end{equation}

En konsekvens är att den relativistiska 'egentiden' $\tau$ (\ref{eq:TAU}) nu kommer att bero på partikelns energi,

\begin{equation}
\label{eq:DSR13}
\Delta \tau = \sqrt{1 - v^2} \, \Delta t \, \frac{f_1({E}_0, \lambda)}{f_1(\tilde E_0, \lambda)} = \sqrt{1 - v^2} \, \Delta t \left(1 + (\gamma - 1) \lambda E_0 \right)^{-1},
\end{equation} 

där $\tilde{E}_0$ betecknar den transformerade (enligt (\ref{eq:DSR4a})-(\ref{eq:DSR4b})) vilo-energin $E_0$. 'Tvilling-paradoxen' i denna version leder till att tidsdifferensen mellan tvillingarna också kommer att bero på deras massa och uppenbarligen går den tyngre tvillingens klocka en aning långsammare. En principiell följd av (\ref{eq:DSR13}) är att två partiklar med olika energier enligt (\ref{eq:DSR11}) kan vara associerade med skilda metriker vid en och samma punkt $(t, x)$ (Magueijo \& Smolin har av denna anledning kallat (\ref{eq:DSR11}) för en 'regnbågsmetrik', 'rainbow metric'). ''Thus, quanta of different energies see different classical geometries'' (M\&S, \url{gr-qc/0305055}). Detta har t ex tolkats som så att varje partikel definierar ett eget momentum-beroende referenssystem som innebär att vid jämförelse mellan två referenssystem måste man förutom den relativa hastigheten (Lorentz) också ta hänsyn till partiklarnas momentum. Dylika överläggningar har föranlett Magueijo och Smolin att hävda att frågan om den 'rätta' duala rumtiden till momentum-rummet ''is the wrong question to ask and that, instead, there is no single classical spacetime geometry when effects of order $L_P$ are taken into account. Instead, we propose that classical spacetime is to leading order in $L_P$ represented by a one parameter family of metrics, parametrized by the ratio $E/E_P$  (...) This may seem nonsensical if one regards the geometry of spacetime as primary. However, in the currently well studied theories of quantum gravity, including loop quantum gravity and string theory, the classical geometry of spacetime is not primary. Rather, it emerges as a low energy coarse grained description of a very different quantum geometry'' (\textit{ibidem}). Detta 'emerges as ...', borde man inflika, är tillsvidare närmast en from förhoppning. Smolin, som är en av de främsta proponenterna för 'loop-gravity', brukar understryka vikten av teorins bakgrundsoberoende; d.v.s., den får inte förutsätta någon \textit{a priori} metrik, och hans mer generella slutsats härav är att teorin inte längre kan leva på någon mångfald ... ''what is left is only algebra, representation theories of algebras, and combinatorics''.\footnote{L. Smolin, ''Quantum theories of gravitation'' i J. D. Barrow et al., 2004, s. 523. Einstein avlsutar också \textit{The meaning of relativity} (1956 uppl.) med anmärkningen att kvantfenomen ''does not seem to be in accordance with a continuum theory, and must lead to an attempt to find purely algebraic theory for the description of reality. But nobody knows how to obtain the basis for such a theory.'' } Som sådan verkar detta vara en något anemisk idé och igen saknar man en vägledande fysikalisk tanke eller princip. Liksom Wheelers dictum ''it all comes from higgledy-piggledy'' förefaller det åtminstone inte tillsvidare ha transmuterats till någon genomgripande fysikteori.

\section{Kvantrum}

Variationerna på temat deformerade Poincaré-algebran är intressanta ur den synvinkeln att ifall en konsistent  kvantgravitation-teori är möjlig så kan DSR-teorier och liknande ge vinkar om diverse sorts approximationer till den 'fullständiga teorin'. En modell som i detta sammanhang verkar ofta återupptäckas av forskare är det veterligen första förslaget till en kvantiserad rumtidsteori som lades fram av Hartland S. Snyder,\footnote{H. S. Snyder, ''Quantized space-time,'' Phys. Rev. 71 (1), 38-41 (1947). Trots efterforskning på internet hittade jag inga närmare biografiska detaljer om Hartland Sweet Snyder (1913-1962). Claus Montonen lyckades gräva fram en kort minnesruna i Physics Today (July 1962). Uppenbarligen dog Snyder i sviterna av en hjärtattack den 22 maj 1962. I Kip S. Thorns \textit{Black holes \& time warps} (W. W. Norton \& Company 1994, s. 212) hittar jag ett kort beskrivning av Snyder som uppenbarligen är baserad på Thornes intervju med Robert Serber, en annan av Oppenheimers elever. ''Snyder was different from Oppenheimer's other students. The others came from middle-class families; Snyder was working class. Berkeley rumor had it that he was a truck driver in Utah before turning physicist. As Robert Serber recalls: 'Hartland pooh-poohed a lot of things that were standard for Oppie's students, like appreciating Bach and Mozart and going to string quartets and liking fine food and liberal politics (...) Of all Oppie's students Hartland was the most independent (...) Hartland had more talent for difficult mathematics than the rest of us', recalls Serber. 'He was very good at improving the cruder calculations that the rest of us did'.'' Förutom svarta hål förefaller Snyder ha varit känd i samband med uppfinnandet av 'strong focusing' tekniken år 1952 för acceleratorer. Snyders kvantrumtidsteori följdes genast upp av C. N. Yang i Phys. Rev. 72, 874 (1947) som inför translationsinvarians. Snyder fick förmodligen uppslaget till teorin från sin lärare Oppenheimer som i sin tur tippsats av W. Pauli som fått idén från W. Heisenberg; nämligen, Heisenberg hade för Pauli framkastat tanken om att införa en osäkerhetsrelation för koordinater. Se F. A. Schaposnik, ''Three lectures on noncommutative field theories'' (\url{hep-th/0408132}). Pauli kommenterar Snyder (1947) i ett brev till Bohr (28.1.1947): ''On the other hand I am looking as critical as you on this idea of so called universal length. If this length - let us call it $l_0$ - is understood to be of geometrical nature, such theories or models will always lead to strange consequences for large momenta of the order $h/l_0$ in a field of purely classical experiments where the quantum of action should not play any role. Recently we discussed here at Zürich a mathematically ingeniuous proposal of Snyder, which, however, seems to be a failure for reasons of physics of the type just mentioned.'' (O. Stoyanov har satt ut brevet på webben, se \url{www.math.uiuc.edu/~stoyanov/math428-M1/}, kopierad från Wolfgang Pauli, \textit{Scientific Correspondence} vol. III, s. 414, ed. Karl von Meyenn, Springer-Verlag, 1985.) Heisenberg intresserade sig också för en sorts gitterteori (lattice theory, Gitterwelt) för rumtiden under 1930-talet. Kanhända p.g.a. den negativa responsen från Bohr så publicerade han aldrig något om teorin även om hans idéer blev kända och diskuterade bland fysikerna (se B. Carazza \& H. Kragh, ''Heisenberg's laittice world: The 1930 theory sketch,'' Am. J. Phys. 63, 7, 595-605, 1995). Bohr ansåg (i ett brev till Mott 18.10.1929) att hypotesen om en minsta  längdenhet i naturen stred mot relativitetsteorin: ''To my view all such limitations would interfere with the beauty and consistency of the theory [of relativity] to far great extent''. Nämligen, gittermodellen utgick från proton Compton-längden som minsta längdenhet och denna är förstås ingen relativistisk invariant. En tid senare (1.4.1930) tillade han dock: ''Although I still think that my arguments were correct, I have revised my attitude towards the matter and think now that very general arguments can be given in favour of such a limitation of space determinations, and that this very point is of essential importance as regards obtaining consistency in the apparent chaos of relativity quantum mechanics.'' (Citaten från Carazza \& Kragh 1995.) Trots en viss enstusiasm övergav Bohr idén om den granulära etermodellen bl.a. eftersom Heisenbergs modell ledde till brott mot laddnings-konservering. Pauli verkar redan från början ha ogillat tanken om minsta längdenheter och varnade att ''those who are making holes in continuous space should mind where they step'' (Pauli under en kongress i Odessa 1930, citerad i Carazza \& Kragh 1995). För en översikt av olika försök att generalisera/modifiera rumtidsbegreppet se N. A. M. Monk, ''Conceptions of space-time: problems and possible solutions,'' Stud. Hist. Phil. Mod. Phys. 28 (1), 1-34 (1997). Monk har samarbetat med Basil Hiley, Bohms nära medarbetare under 80-talet, som det senaste årtiondet försökt utveckla en sorts algebraisk teori för fysiken och intresserat sig för 'icke-kommutativ kvantgeometri' (se Hileys bidrag i \textit{Quo vadis quantum mechanics?} A. Elitzur, S. Dolev \& N. Kolenda, eds. (Springer, 2005) s. 299-324). C. F. v. Weizsäcker, en elev till Heisenberg, har under en lång era arbetat på en sorts kvantlogiskt fundament ('Urtheorie') för fysiken i ett försök att s.a.s. förena Einstein, Bohr och Heisenberg (se \textit{Aufabau der Physik} (Carl Hanser Verlag, 1985)). En intressant omständighet är att Weizsäcker (\textit{Aufbau}, kap.9, ''Spezielle Relativitätstheorie'') kommer fram till gruppen $SO(2,4)$ som också centralt figurerar i R. Penroses twistorteori. Penrose har i sin tur i över 40 år spunnit på med sina twistorer men i sitt epos \textit{The road to reality} (Jonathan Cape, 2004) hamnar han tillstå att teorin tillsvidare främst har gett en alternativ formulering av kända resultat inom relativistisk fysik. Han betonar att twistor-teorin inte är en fysikalisk teori, men Penroses förhoppning är dock att den skall bana väg för ny fysik i likhet med Hamilton-formalismen i klassisk fysik. ''Hamiltonian theory did not introduce physical changes, but it provided a different outlook on classical physics that later proved to be just what was required for the new quantum theory''... (\textit{ibid.} s. 1004).} elev till J. R. Oppenheimer (Snyder och Oppenheimer var f.ö. de första att presentera ett scenario för hur ett svart hål bildas genom en implosion av en tryckfri vätska i ett gemensamt arbete i Phys. Rev. 56, 455-459, 1939). Under 30- och 40-talet inspirerades arbeten kring rumtids-kvantisering av tanken att man på detta sätta skulle kunna åtgärda divergensproblemen i kvantfältteorierna. ''We hope that the introduction of such a unit of length [parametern $a$ nedan] will remove many of the divergence troubles in present field theory'', anmärker Snyder i inledningen till sin artikel. I och med att renormaliseringstekniken utvecklades (Feynman, Dyson, Tomonaga, Schwinger, Bethe, m fl) bortföll denna motivering och arbeten på rumdtidskvantiseringar stannande mer eller mindre av. Divergensproblemen inom kvantgravitation och utveckligen av icke-kommutativ geometri har emellertid återupplivat intresset för dylika teorier (vore intressant att följa upp hur citationskurvan för Snyders artikel utvecklats med tiden).\footnote{Snyders papper varit mer eller mindre bortglömd tills nyligen då den återupptäckts i samband med utvecklingen av icke-kommutativ geometri, kvantgrupper och DSR. Man skulle kanske tycka att kvantrumtid hade intresserat dem som studerat relationer mellan kvantteori och gravitationen. Även om t ex C. W. Misner, K. S. Thorne \& J. A. Wheeler i \textit{Gravitation} (Freeman, 1973) spekulerar om pregeometri, 'space-time foam', etc, nämns inte Snyder. Två färska böcker i kvantgravitation, C. Rovelli (2004) och C. Kiefer (\textit{Quantum gravity} (Oxford U. P., 2004)), nämner varkendera kvantrumtid eller Snyder. Första gången jag själv stötte på en hänvisning till Snyder (1947) var i von Weizsäckers \textit{Aufbau} (1985). En av de få böcker överhuvudtaget som varit dedikerad åt ämnet rumdtid och kvantmekanik torde vara D. I. Blokhintsev, \textit{Space and time in the microworld} (Reidel, 1973).} En tämligen direkt procedur för att erhålla en icke-kommutativ geometri är att tolka koordinaterna $x^i$ som rotationsoperatorer i ett utvidgat rum. Detta är vad Snyders gör genom att uppfatta koordinaterna som hermitska operatorer, vars egenvärden motsvarar de mätbara storheterna (''possible results of measurements''), genom att utgå från den reella kvadratiska formen

\begin{equation}
\label{eq:QST1}
-\eta^2 = \eta_0^2 - \eta_1^2 -\eta_2^2 -\eta_3^2 -\eta_4^2 \,, 
\end{equation}

som för konstant $\eta$ beskriver ett 4-dimensionellt de Sitter-rum. Koordinaterna $x^{\mu}$ tolkas då som differentialoperatorer av formen (användande metriken $h = \mbox{diag} (1,-1,-1,-1,-1)$ och $\eta_{\alpha} = h_{\alpha \beta} \eta^{\beta}$)

\begin{equation}
\label{eq:QST2}
x^{\mu} = i a \left( \eta^{\mu}\frac{\partial}{\partial \eta_4} -\eta^{4}\frac{\partial}{\partial \eta_{\mu}}  \right)
\quad (\mu = 0, 1, 2, 3), 
\end{equation}

med $a$ som ''the natural unit of length''. För dessa operatorer får vi

\begin{eqnarray}
\label{eq:QST3a}
[x^{\mu}, x^{\nu}] &=& \frac{i a^2}{\hbar} L^{\mu \nu} \quad \mbox{med}\\
\label{eq:QST3b}
L^{\mu \nu} &\equiv& i \hbar \left( \eta^{\mu}\frac{\partial}{\partial \eta_{\nu}} -\eta^{\nu}\frac{\partial}{\partial \eta_{\mu}}    \right).
\end{eqnarray}

För $\mu$ = 1, 2, 3 har $x^{\mu}$ således formen av en rotationsoperator med egenfunktioner av formen $e^{im\phi}$ som därför implicerar egenvärdena $ma$, för $m \in \mathbf{Z}$; d.v.s., multipler av längdenheten $a$. För tidsoperatorn $x^0$ däremot har vi en 'boost'-typ operator som inte leder till någon motsvarande 'kvantisering' av tiden. Att 'tiden' här tack vare pseudometriken istället får ett kontinuerligt spektrum kan måhända antyda en poäng med att 'tiden' skiljs ut med sitt speciella förtecken i metriken. Kvantiseringen av rumtiden verkar således implicera en asymmetri, diskret vs kontinuerlig, mellan rum och tid som inte direkt återfinns i SR. För impulsvariablerna valde Snyder (ett av många matematiskt rimliga alternativ) formen

\begin{equation}
\label{eq:QST4}
p^{\mu} = \frac{\hbar}{a} \frac{\eta^{\mu}}{\eta^4},  
\end{equation} 

vilket är förenligt med den konventionella relationen $L^{\mu \nu} = x^{\mu} p^{\nu} - x^{\nu} p^{\mu}$ och ekv (\ref{eq:QST3b}). Från (\ref{eq:QST4}) följer t ex kommuteringsrelationen

\begin{equation}
\label{eq:QST5}
[x^1, p_1] = i \hbar \left\{1 + \left(\frac{a}{\hbar}\right)^2 p_1^2 \right\}.  
\end{equation}

Ekv (\ref{eq:QST5}) implicerar i sin tur en osäkerhetsrelation av typen (som numera går under beteckningen 'generalized uncertainty relation/principle', GU)

\begin{equation}
\label{eq:QST6}
\Delta x \Delta p \geq \frac{\hbar}{2} \left\{1 + \left(\frac{a}{\hbar}\right)^2 {\langle p^2\rangle} \right\},  
\end{equation}

som bl a aktualiserats inom strängteorin.\footnote{M. Maggiore, ''Quantum groups, gravity, and the generalized uncertainty principle,'' Phys. Rev. D 49, 10, 5182-7, 1994; L. J. Garay, ''Quantum gravity and minimum length,'' Int. J. Mod. Phys. A10, 145, 1995 (\url{gr-qc/9403008}).} (Mer exakt borde vi för (\ref{eq:QST6}) ha tagit kvadratroten av medelvärdet av kvadraten av högra ledet i (\ref{eq:QST5}), $\sqrt{\langle ( \dots )^2 \rangle}$.) Ekv (\ref{eq:QST6}) innebär vidare att 

\begin{equation}
\label{eq:QST7}
\Delta x \approx \frac{\hbar}{2} \left\{\frac{1}{\Delta p} + \left(\frac{a}{\hbar}\right)^2 \Delta p \right\}.  
\end{equation}

Minimerar vi högra delen i (\ref{eq:QST7}) visavis $\Delta p$ får vi olikheten $\Delta x >  a$ (minimivärdet motsvarar $\Delta p = \hbar/a$) vilket åter visar att $a$ karaktäriserar en sorts minsta mätbara längdenhet hos rummet.

I en uppföljande artikel\footnote{H. S. Snyder, ''The electromagnetic field in quantized space-time,'' Phys. Rev. 72, 1, 68-71, 1947. } formulerade Snyder Maxwells ekvationer i denna modell för kvantrumtid och antydde att Diracs, Procas och Kleins ekvationer kunde behandlas analogt, men detta projekt förefaller ha avbrutits. (Snyders sista vetenskapliga publikation i \textit{Physical Review} är från år 1955.) Sju år senare publicerar Hellund och Tanaka en artikel\footnote{E. J. Hellund \& K. Tanaka, ''Quantized space-time,'' Phys. Rev. 94, 1, 192-195, 1954. H\&T-modellen är en sorts reciprok variant av Snyder-modellen.} baserad på en variation av Snyders modell där de bl a uppmärksammar relationen (\ref{eq:QST6}) och undersöker Dirac-ekvationen och möjligheten av lösningar som ger 'time-localized states'. De senaste åren kan utvecklingen sägs ha fortsatt i form av fält-teori på icke-kommutativa rum.\footnote{För översikter se Schaposnik (\url{hep-th/0408132}); A. Tureanu, ''Some aspects of quantum field and gauge theories on noncommutative space-time,'' Helsingfors Universitet 2004, Ph.D.-avhandling, e-version (pdf): \url{http://ethesis.helsinki.fi/julkaisut/mat/fysik/vk/tureanu/}.} Ett enkelt sätt att införa icke-kommutativitet är att för klassiska kommutativa koordinater $x^i$ definiera koordinatoperatorer

\begin{equation}
\label{eq:QST8}
\hat{x}^i = x^i - \frac{1}{2 \hbar} \Theta^{ij} p_j,  
\end{equation}

där $p_j = - i \hbar \partial / \partial x^j$  och $\Theta$ är en konstant antisymmetrisk matris. Av (\ref{eq:QST8}) följer att

\begin{equation}
\label{eq:QST9}
[\hat{x}^i , \hat{x}^j] = \Theta^{ij}.  
\end{equation}

Calmet (2005)\footnote{X. Calmet, ''Space-time symmetries of noncommutative spaces,'' Phys. Rev. D 71, 085012, 2005.} argumenterar i omvänd riktning; givet (\ref{eq:QST9}) så kan vi införa kommuterande koordinater $x^i$ via (\ref{eq:QST8}).

\section{Avslutning}

I detta skede, mer än 100 år efter publiceringen av SR, kan vi konstatera att  Relativitetsteorin är mer befäst än någonsin. Spekulationer om avvikelser från SR har förblivit blott spekulationer. Som antytts återstår dock många fundamentala problem inom fysiken som kan ha konsekvenser för SR. På det teoretiska området kan emellertid skönjas en sorts utmattning.\footnote{Matematikern-fysikern Peter Woit har i sin \textit{Not even wrong. The failure of string theory and the continuing challenge to unify the laws of physics} (Jonathan Cape, 2006) gått till attack mot sträng-teorin eftersom den inte har lett till en enda verifierbar förutsägelse och menar att den mer liknar religion än vetenskap. De mer begränsade ansatserna till kvant-gravitation har ej heller lett till något genombrott. Lee Smolin som jobbat med loop-kvantisering verkar i \textit{The trouble with physics: The rise of string theory, the fall of a science, and what comes next} (Houghton Mifflin, 2006) också ganska desperat och efterlyser en ny Einstein. Smolins bok pryds förresten av en pärmbild visande ett par skor med hopbundna skosnören ... } 'Sträng-teorins' 20-åriga ökenvandring har inte verkat föra oss närmar det förlovade landet som skulle bl a klara upp kvant-gravitationen. Men liksom tidigare under historiens gång kan den förlösande insikten komma närsomhelst från ett oväntat håll.

\end{document}